\documentclass[fleqn,10pt]{wlscirep}
\usepackage[utf8]{inputenc}
\usepackage[T1]{fontenc}
\usepackage{bm}
\title{Predicting potato plant vigor from the seed tuber properties}

\author[1,*]{Elisa Atza}
\author[2]{Rob Klooster}
\author[2]{Falko Hofstra}
\author[2]{Frank van der Werff}
\author[2]{Hans van Doorn}
\author[1,**]{Neil Budko}
\affil[1]{Numerical Analysis, Delft Institute of Applied Mathematics, Faculty of Electrical Engineering, Mathematics and Computer Science, Delft University of Technology, Mekelweg 4, 2628 CD Delft, Netherlands}
\affil[2]{HZPC Research B.V., Roptawei 4, 9123 JB Metslawier, Netherlands}
\affil[*]{e.atza@tudelft.nl}
\affil[**]{n.v.budko@tudelft.nl}

\begin{abstract}
The vigor of potato plants, defined as the canopy area at the end of the exponential growth stage, depends on the origin and physiological state of the seed tuber. Experiments carried out with six potato varieties in three test fields over three years show that there is a 73\%-90\% correlation in the vigor of the plants from the same seedlot grown in different test fields. However, these correlations are not always observed on the level of individual varieties and vanish or become negative when the seed tubers and young plants experience environmental stress. A comprehensive study of the association between the vigor and the seed tuber biochemistry has revealed that, while 50\%-70\% of the variation in the plant vigor is explained by the tuber data, the vigor is dominated by the potato genotype. Analysis of individual predictors, such as the abundance of a particular metabolite, indicates that the vigor enhancing properties of the seed tubers differ between genotypes. Variety-specific models show that, for some varieties, up to 30\% of the vigor variation within the variety is explained by and can be predicted from the tuber biochemistry, whereas, for other varieties, the association between the tuber composition and the vigor is much weaker.

\end{abstract}
\begin{document}

\flushbottom
\maketitle

\thispagestyle{empty}

\section*{Introduction}

Potato ({\it Solanum tuberosum}) is a nutritious staple food, popular in many countries of the world, and an important source of starch and proteins. According to the Food and Agriculture Organization of the United Nations, in 2022, almost 375 million metric tonnes of potato were produced globally with the estimated value of 120 billion USD \cite{FAOSTAT-database}. While potato production is a multi-billion dollar industry, the profit margins of the potato seed producers are relatively low compared to other popular crops, such as tomatoes, onions or lettuce. The reason is that potatoes are not grown from the seedlings found inside the berries of the potato plants but rather from other potatoes, called seed tubers. These seed tubers are cultivated separately, in clean soil that is free from diseases, and stored in controlled conditions that prevent premature sprouting, loss of moisture, rotting and physiological disintegration of the tubers \cite{Seed-potato-book-2023}. This relatively expensive and complicated production cycle is the main reason why some of the technologies common in the seed-based crops are still not available in the potatoes industry. In particular, there is no reliable germination test for potatoes, such as the one, e.g., for barley, where the vigor of a seedlot is ascertained by letting some seeds germinate on a piece of wet paper.

Over the years, producers of the seed potatoes have accumulated qualitative evidence about the lack of consistency in the performance of their produce. This lack of consistency manifests itself as a noticeable difference in the size of the emergent potato plants of the same variety, where some plants appeared to be less `vital' or `vigorous' than others. Since the seeds of the same variety are exact clones, this difference could not be attributed to the genetic factors and was traced back to the difference in the production field or even a farmer who might have used a slightly different crop management technique. This paper reports on the results of the project `Flight to Vitality' (FtV) which was the result of cooperation between an EU funding agency, two major seed producers, and two academic institutions in the Netherlands. The main idea of the FtV project was to study the relation between the properties of the seed tubers and the vigor of the resulting potato plants. The ultimate goal was to build a data-driven model that would help the seed producers and the farmers to rank the seedlots of different production origin by their quality.

\begin{figure}
    \centering
    \begin{minipage}[c]{0.55\linewidth}
    \begin{minipage}[c]{0.75\linewidth}
                    \includegraphics[width=0.97\linewidth , trim={6.8cm  0 0 0 }, clip]{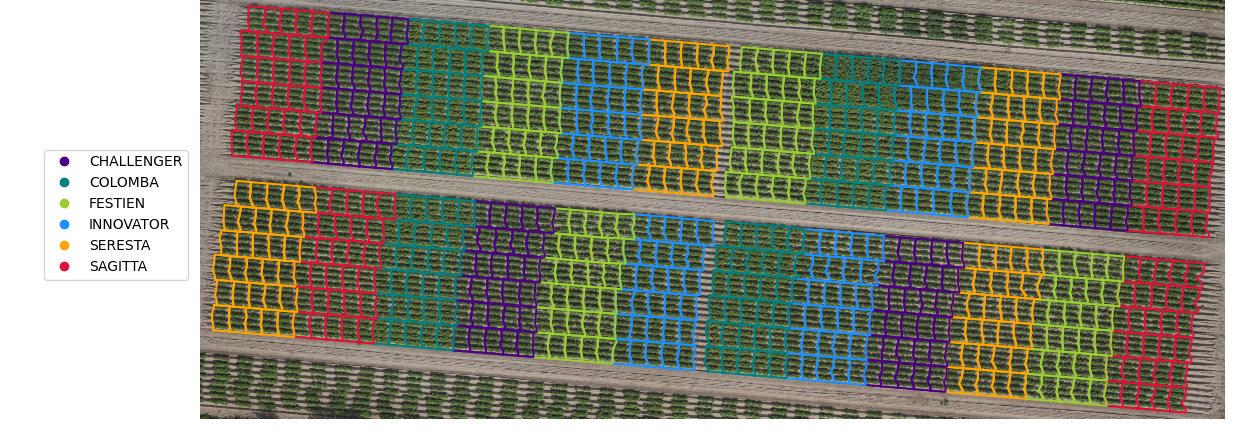}   
    \end{minipage}
        \begin{minipage}[c]{0.22\linewidth}
                    \includegraphics[width=0.99\linewidth, trim={1.55cm 5cm 37.2cm 5cm},clip ]{Images/Introduction/Ass_V_2021_h.png}
    \end{minipage}
  \vfill  \includegraphics[width=\linewidth, trim={0 0 5.5cm 0}, clip]{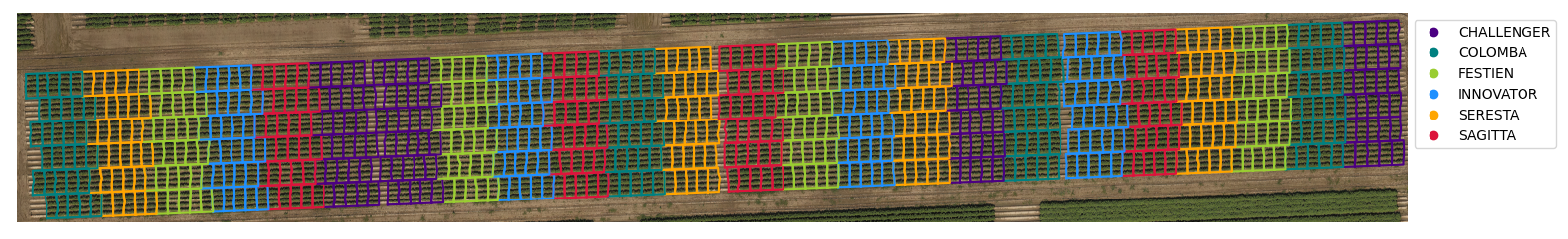}

    \caption{Drone images of the V (top) and S (bottom) test fields in 2021 with the overlayed plot boundaries. Colors correspond to the six varieties planted in a randomized block design.}

    \label{fig:field_experiment_design}
    \end{minipage}
    \begin{minipage}[c]{0.44\linewidth}
        \includegraphics[width=0.98\linewidth]{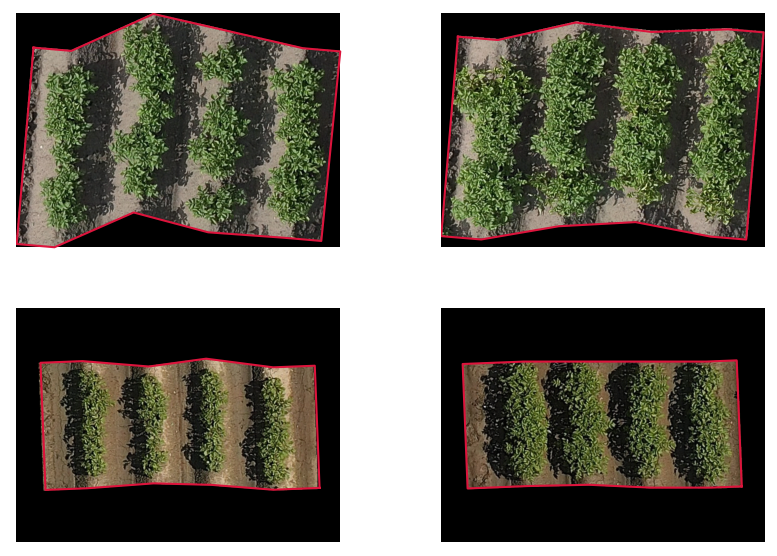}
        \caption{Plots of weak (left) and strong (right) seedlots of the Sagitta variety showing consistent performance between the V (top) and the S (bottom) test fields.}
    \label{fig:plots_images}
    \end{minipage}

\end{figure}

In this project, six potato varieties (Challenger, Colomba, Festien, Innovator, Sagitta, and Seresta) have been studied, each represented by 30 seedlots of different production origin. The tubers from these seedlots were planted in three test fields, utilizing the randomized spatial design, see Figure~\ref{fig:field_experiment_design}. The test fields were located in Montfrin (France), Veenklooster (Netherlands), and Kolummerwaard (Netherlands, SPNA), and will be denoted as the fields M, V, and S, respectively. The experiment, including the selection of the new 30 seedlots of each variety, biochemical and microbial analysis of the tubers, and the subsequent planting and observation in the three test fields, was repeated over three consecutive years (2019, 2020, and 2021). 

The size and health of the leaf canopy are the most readily observable features of potato plants. In fact, observations of the difference in the canopy size have lead to the qualitative conclusions about the differences in the seed tuber vigor. These features are also relatively easy to monitor with a drone-mounted camera. Finally, the size of the leaf canopy at the time of tuber bulking is indicative of the eventual yield \cite{canopy-yield}. Therefore, the canopy area was chosen as the vigor parameter in this project (see Methods). This choice also agrees with the accepted definition of the potato seed tuber's vigor\cite{Yang-et-al-2024,Potato-vigor}. In Figure~\ref{fig:plots_images}, one can see the example images of two Sagitta seedlots showing systematic difference in the canopy size on the chosen date in two test fields.

In our previous publication \cite{Yang-et-al-2024} which included partial results of the FtV project, the relation between the microbiome and the vigor of the plants has been considered, and the ability of microbiome to explain up to 20\% of the vigor variation within some varieties in the tests run over two consecutive years (2019 and 2020) has been demonstrated. There is certainly an intimate connection between the microbiome found on the skin and in the `eyes' of the seed tuber and the tuber biochemistry. Some fungal and viral species may also directly affect the growth and development of the potato leaf canopy, e.g., by inducing diseases such as early or late blight, leaf roll, verticillium wilt, etc. \cite{potato-diseases}  However, the presence and abundance of microbiome species on the surface of the tubers is to a large extent determined by the field of production (soil type, treatment, etc.) and represents only an indirect measure of the tuber biochemistry, which is known to affect the early growth stages \cite{tuber-reserves}. In the present paper the focus is on the actual biochemical properties of the seed tubers as provided by the four standard laboratory techniques: X-Ray Fluorescence (XRF), Fourier-Transform Infrared Spectroscopy (FTIR), Hyper-Spectral Imaging (HSI), and High-Resolution Mass Spectroscopy (HRMS, untargeted metabolome). These techniques quantify, respectively: the abundance of inorganic elements, the far-infrared spectra pertaining to organic compounds in the dry and wet samples, and the abundance of all detectable small molecules (metabolites).

It is well known that the quality of predictions may depend on the choice of the mathematical model that seeks to connect the predictor variables with the dependent variable. It the present case the number of predictors (biochemical information about the tubers) significantly exceeds the number of experiments (measurements of the vigor parameter) -- the situation known as overparameterization or $p\gg n$ in the Machine Learning context. 
Nevertheless it has been shown \cite{ANNvsMLR2022, DNNvsLR2020} that the use of complex models does not result in a performance improvement for datasets of limited size such as ours. 
In other words, there is no reason to expect a multilayered Artificial Neural Network (ANN), typically applied to overparameterized problems, to perform any better than a simple Multiple Linear Regression (MLR) model. Hence, while in our previous paper \cite{Yang-et-al-2024} a nonlinear Random Forest (RF) model was considered, in this paper we turn to an MLR model, common in the FTIR and HSI applications \cite{ChemometMLR2001,Chemomet2005,Chemomet2023}. Specifically, a version of the fixed-effect MLR model called the Inverse Regression (IR) model \cite{Atza-Budko-2024} is applied. The IR model is especially suited for severely overparameterized problems with diverse predictor data and has been further adapted here to accommodate for the possible discontinuities in the dependence of the tuber properties on the plant vigor (see Methods). The IR method also allows to compare the quality of the predictor data and to remove the features (predictors) that are either too noisy or do not comply with the linearity assumptions of the MLR model.

\section*{Results}

\subsection*{Consistency of the seedlot vigor}

\begin{figure}
    \centering
                \includegraphics[width=0.5\linewidth]{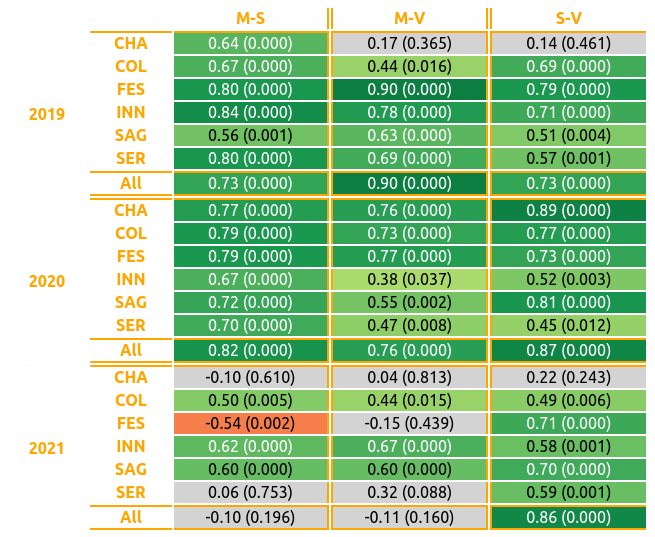}
    \caption{Observed correlations (Pearson) in the canopy area on selected dates (see Methods and Table~\ref{tab:choice_dap}) between the three fields (M, S, and V) in each year. The rows show the correlations for each variety as well as for all varieties together ('All').
    Dark-green background highlights large significant positive correlations, red background indicates the significant negative correlation. The corresponding p-values are displayed between the brackets, insignificant correlations, i.e., with the p-value larger than $0.05$, are displayed on a gray background.}
    \label{fig:correlations}

\end{figure}

Since each year the seed tubers from the same seedlot were planted in three test fields at different locations and times, it was possible to investigate the consistency of the seedlot vigor (canopy cover area at a specific date after planting, see Methods) in a quantitative way. The Pearson correlation coefficients between the vigor data vectors are shown in Figure~\ref{fig:correlations}. There is a consistent positive significant correlation between the full vigor vectors (all varieties) in most of the fields, see the row marked as "All". The results for the first two years (2019 and 2020) of the experiment were already presented in our previous publication about this project \cite{Yang-et-al-2024}. Here the third year of the experiment (2021) has been included as well and the canopy measurement algorithm has been improved for all three years (see Methods). 

The second and especially the third year of the project have demonstrated that, while the correlations for the individual varieties are often strong, positive and significant, they can sometimes be affected by environmental conditions and external events. Specifically, in the test field V in 2020, the lack of rain around the emergence time has caused the remaining herbicide to hinder the growth of early-emerging varieties and seedlots, which otherwise would have exhibited a strong vigor. In the test field M in 2021, the untypically low temperatures and strong winds around the planting time have delayed the emergence and affected the growth and development of the early-emerging varieties and seedlots. This latter event has resulted in insignificant negative correlations between the field M and the other two fields in 2021 when all varieties were considered together Figure~\ref{fig:correlations} (bottom row), and in the significant negative correlation for the Festien variety (red entry).

The observed canopy size correlations confirm and quantify the qualitative evidence from the potato farming practice that the seed tubers of the same variety originating from different production fields indeed produce plants of consistently different vigor. It is important to note that the 30 seedlots of each variety were exact clones. Hence, the systematic differences in the vigor cannot be explained by the genetic factors only. To avoid the seed aging effect, the consistency of the difference between the seedlots was only tested within a single year following the year of production, i.e., it has been decided not to plant the seeds in the years after that.

It is also clear that not all varieties show the same level of correlation between the test fields. One variety, namely, Festien, stands out as having the most consistent high correlations (70\%-90\%) when the M field in the year 2021 is excluded (this test field also shows the most significant negative correlation with respect to other fields for this variety). While we also observe high correlations for other varieties, these are not as consistent. Note that only the correlations above 50\% result in useful predictions.

\subsection*{Genotype dominates the seedlot vigor}

The present experiments were not designed to recover the genotype-by-environment interaction, which was also not known {\it a priori} for the selected set of varieties. Therefore, the available environmental information (solar radiation, temperature, water etc.) could not be used to its full potential, i.e., in a way that would improve the prediction quality within the specific growth conditions. In these circumstances one can hope to predict the average vigor of a seedlot over several conditions, i.e., over several fields within the same test year. However, this strategy only works if the conditions are not too different between the fields. Since the V field in 2020 and the M field in 2021 did suffer from adverse growth conditions around the emergence times, these fields are considered to be statistical outliers and the corresponding data are treated with extra caution.

\begin{figure}
    \centering
    \includegraphics[width=0.98\linewidth, trim={0 0.7cm  0 0 },clip]{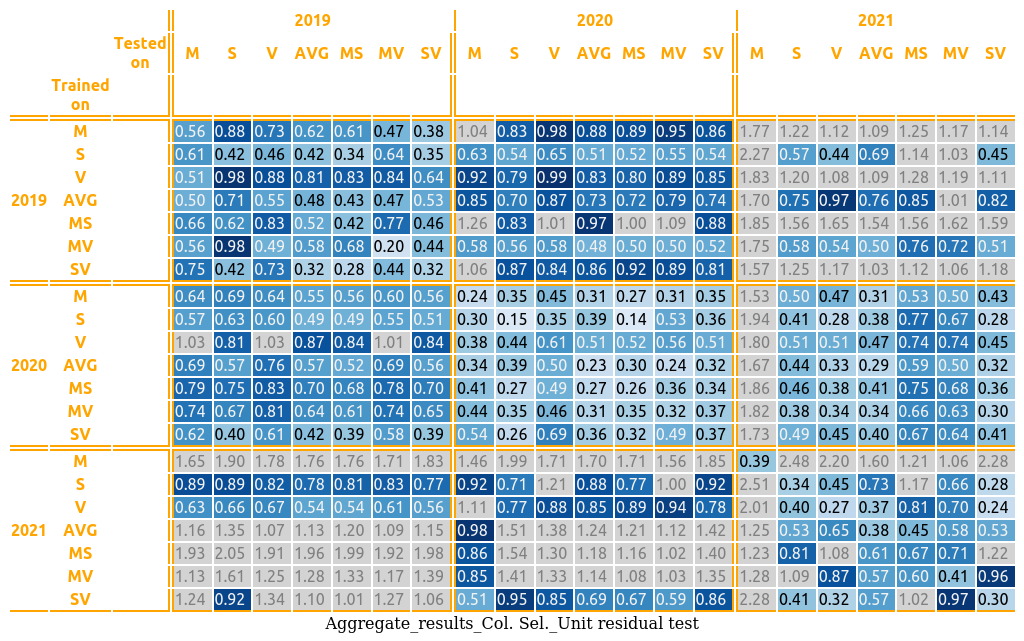}
        \includegraphics[width=0.98\linewidth, trim={0 0.7cm  0 0 },clip]{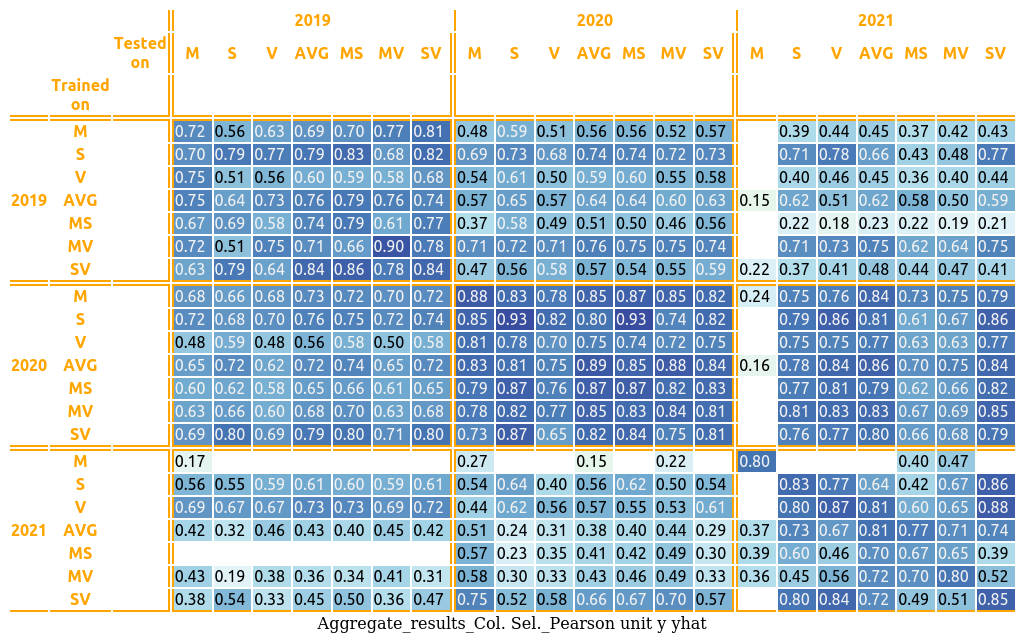}
    \caption{Performance of the variety-agnostic model tested on all six varieties simultaneously: the SSE residual $r^{2}$ (top), the Pearson correlation coefficient $c$ (bottom). Rows correspond to the training dataset, columns show the testing configuration. SSE residuals $r^{2}\geq 1$ are grayed out, correlations with $p>0.05$ are omitted.}
    \label{fig:var_agn_full_field}
\end{figure}

To predict the vigor of a plant from the seed tuber properties, a model should be trained on both the tuber and the vigor data and then a prediction should be made using only the tuber data with the measured vigor data used for the testing purposes only. Since the present experiment spanned three years and three fields in each year, there are many possibilities to train and to test the model. For completeness, all possible training-testing possibilities have been implemented and the results are shown as tables in Figures~\ref{fig:var_agn_full_field}--\ref{fig:SER_pearson_corr_var_agn_var_spec}, where the rows correspond to the training dataset and the columns show the outcome of the testing. Note that the block-diagonal portions of the table correspond to the testing on a randomly selected portion of the dataset that was excluded from the training process. This means that the testing within the same year (diagonal blocks) was performed on a smaller dataset than across the years (off-diagonal blocks) and is therefore less reliable.

The results presented in Figure~\ref{fig:var_agn_full_field} correspond to the variety-agnostic model which does not code explicitly for the known variety of the seedlot, and both the SSE residual or the unexplained variance (top table) and the correlation of the predicted and measured vigor (bottom table) are shown. It turns out that the data from 2020 are most suited for the training of the linear model as it results in the best predictions (smallest residuals and highest correlations) for the years 2019 and 2020. Hence, the variety-agnostic model is able to explain around 70\% of the variation of the vigor of the plants (within the same year) and to predict from 50\% to 70\% of the variation in another year. These residuals correspond to 70\%-80\% correlations between the predicted and measured vitalities, which agrees with the magnitude of the observed correlations (see Figure~\ref{fig:correlations}, rows marked 'All') in vigor across the fields in each year. 

However, these promising regression results mask the fact that the average vigor (canopy size) of the six considered varieties is markedly different, with the canopy of the `weakest' variety Festien (around $-1$ on the normalized field-centered scale) being almost three times smaller than that of the `strongest' variety Colomba (around $2$ on the normalized field-centered scale). Therefore, mathematically speaking, it is easy to achieve a relatively small residual by simply predicting the average vigor of each variety rather than the true vigor of the seedlot.

\begin{figure}
    \centering
    \includegraphics[width=0.98\linewidth, trim={1cm 0 1cm 3cm}, clip]{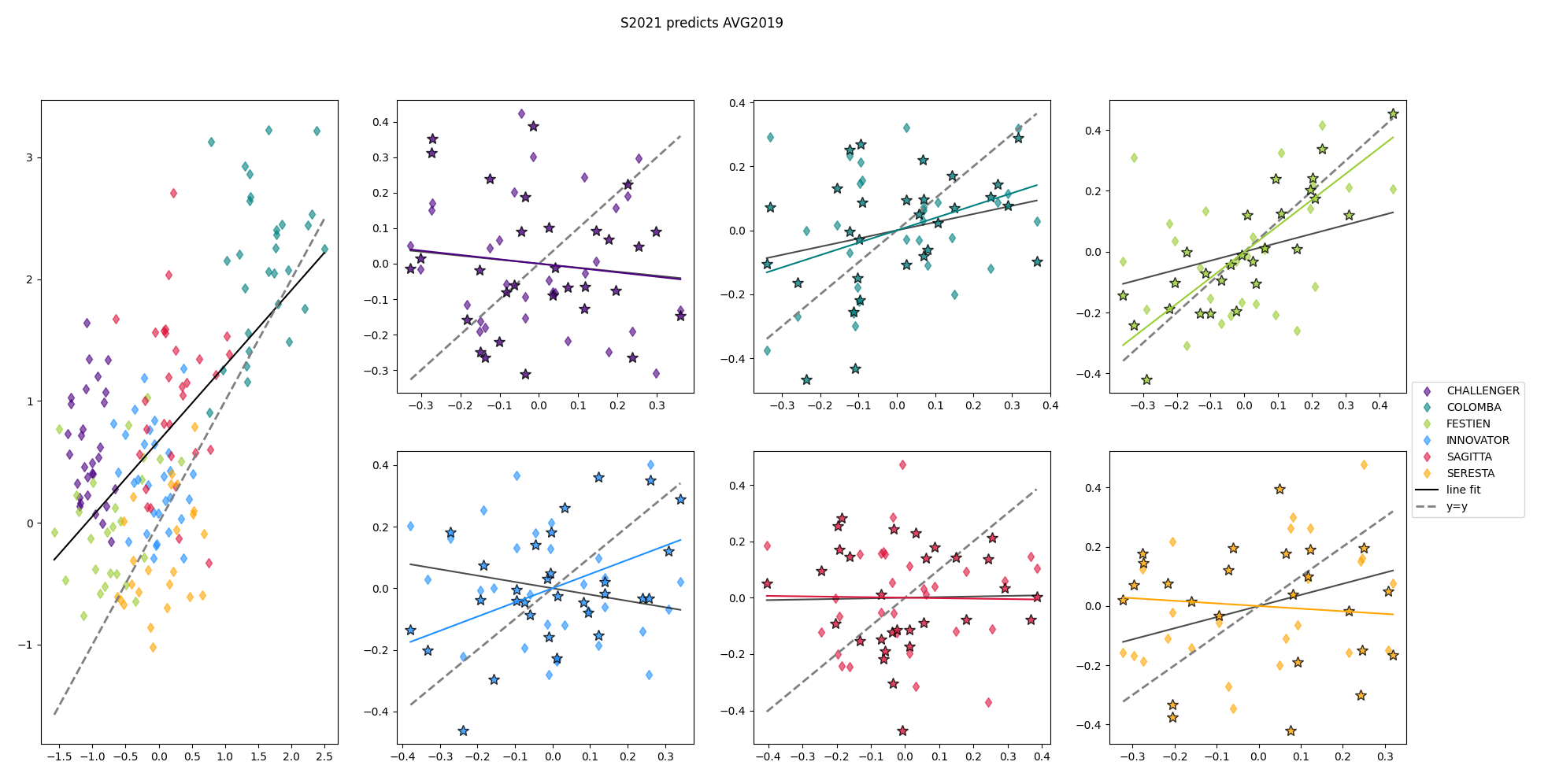}
    \caption{Performance of the variety-agnostic and variety-specific models trained on the dataset of the field S in 2021 and tested on the average vigor data of the fields M, S, and V in 2019. Horizontal axis -- measured seedlot vigor $y$, vertical axis -- predicted seedlot vigor $\hat{y}$.
    Leftmost vertical panel: variety-agnostic model with the seedlot results colored according to their variety. Dashed gray: the $y=\hat{y}$ line; solid black: the best linear fit between the measured and predicted data. Six panels on the right: prediction per variety with the variety-agnostic model (diamonds) and the variety-specific model (stars) with the variety vigor data (both measured and predicted) shifted and scaled to have zero mean and unit norm. Dashed gray: the $y=\hat{y}$ line; solid black: the best linear fit for the variety agnostic model; solid colored: the best linear fit for the variety-specific model.}
    \label{fig:Scatterplot_agn_spec}
\end{figure}

The fact that the genotype (variety) effect dominates the regression results can be deduced from the scattered plots of Figure~\ref{fig:Scatterplot_agn_spec}, where the vigor prediction of each seedlot is colored according to the corresponding genotype. It is obvious that, with some exceptions, the quality of predictions of the seedlot vigor withing each individual variety is rather poor.

\subsection*{Some varieties are more predictable than others}

\begin{figure}
    \centering

        \includegraphics[width=0.98\linewidth, trim={0 0.7cm  0 0 },clip]{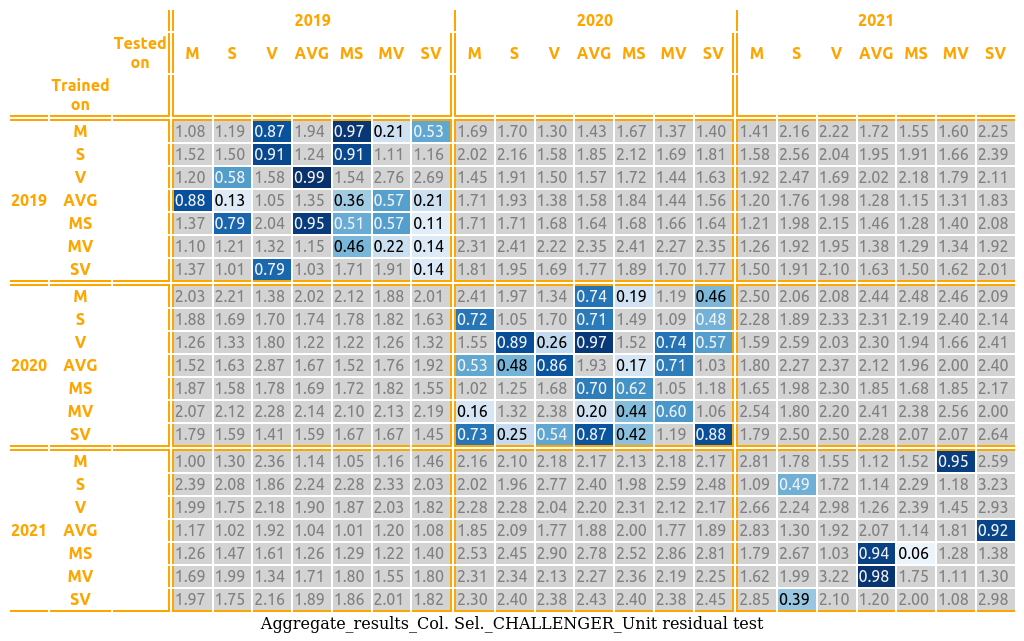}
            \includegraphics[width=0.98\linewidth, trim={0 0.7cm  0 0 },clip]{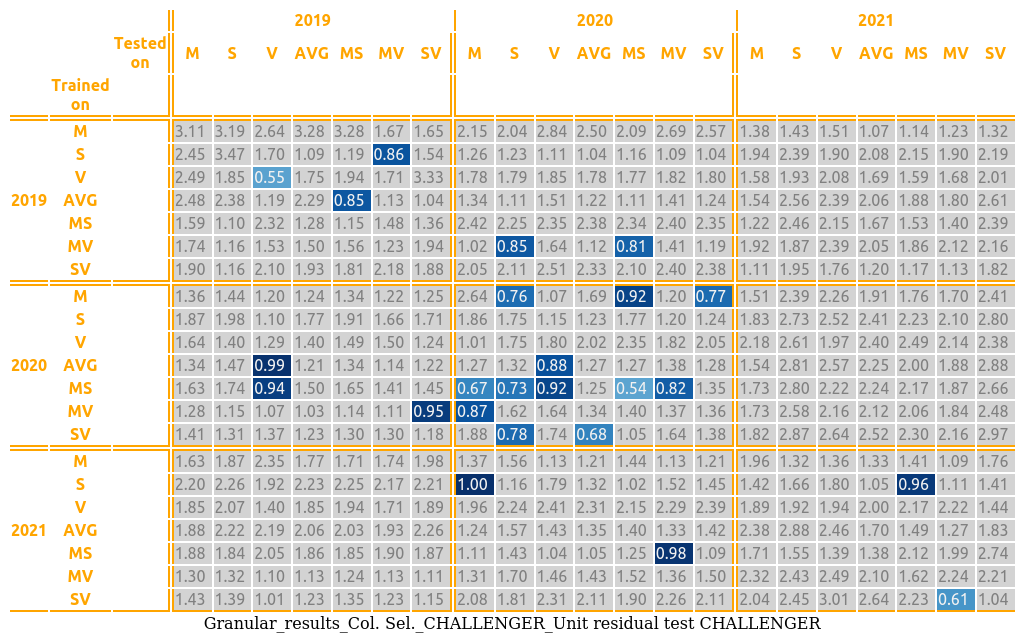}
    \caption{Performance (SSE residual $r^{2}$) of the variety-specific Challenger model (top) and the variety-agnostic model tested on the Challenger variety (bottom). Rows correspond to the training dataset, columns show the testing configuration. SSE residuals $r^{2}\geq 1$ are grayed out.}
    \label{fig:CHA_residuals_var_agn_var_spec}
\end{figure}

\begin{figure}
    \centering

        \includegraphics[width=0.98\linewidth, trim={0 0.7cm  0 0 },clip]{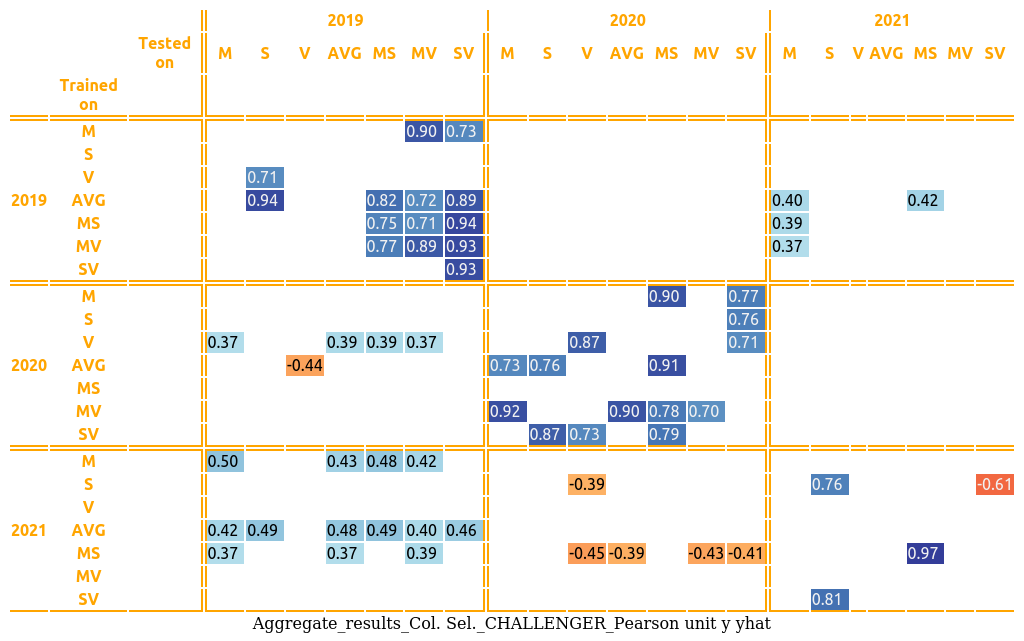}
            \includegraphics[width=0.98\linewidth, trim={0 0.7cm  0 0 },clip]{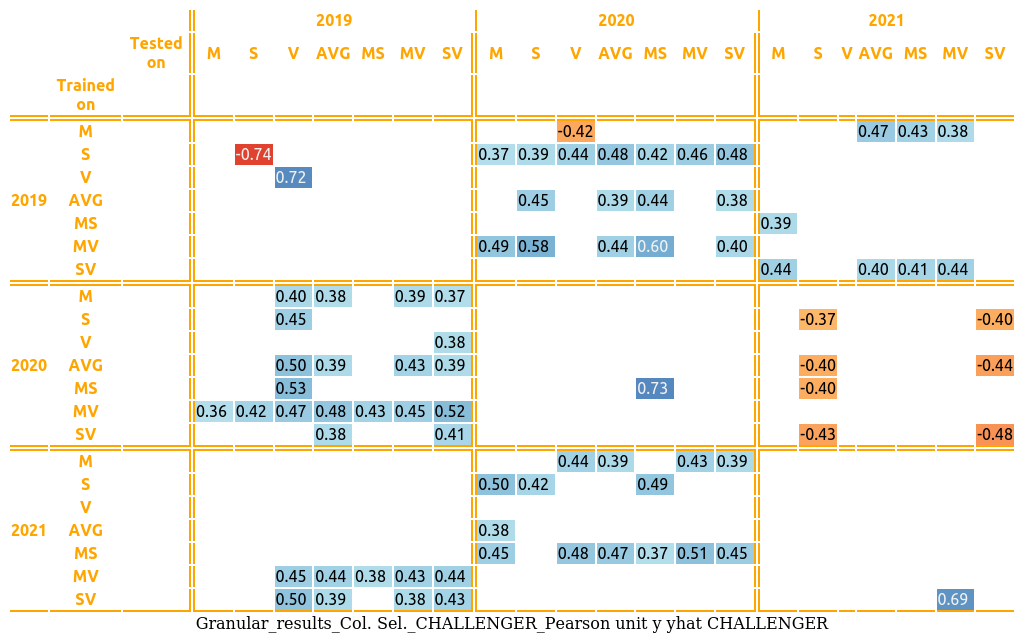}
    \caption{Performance (Pearson correlation $c$) of the variety-specific Challenger model (top) and the variety-agnostic model tested on the Challenger variety (bottom). Rows correspond to the training dataset, columns show the testing configuration. Correlations with $p\geq 0.05$ are omitted.}
    \label{fig:CHA_pearson_corr_var_agn_var_spec}
\end{figure}

\begin{figure}
    \centering

        \includegraphics[width=0.98\linewidth, trim={0 0.7cm  0 0 },clip]{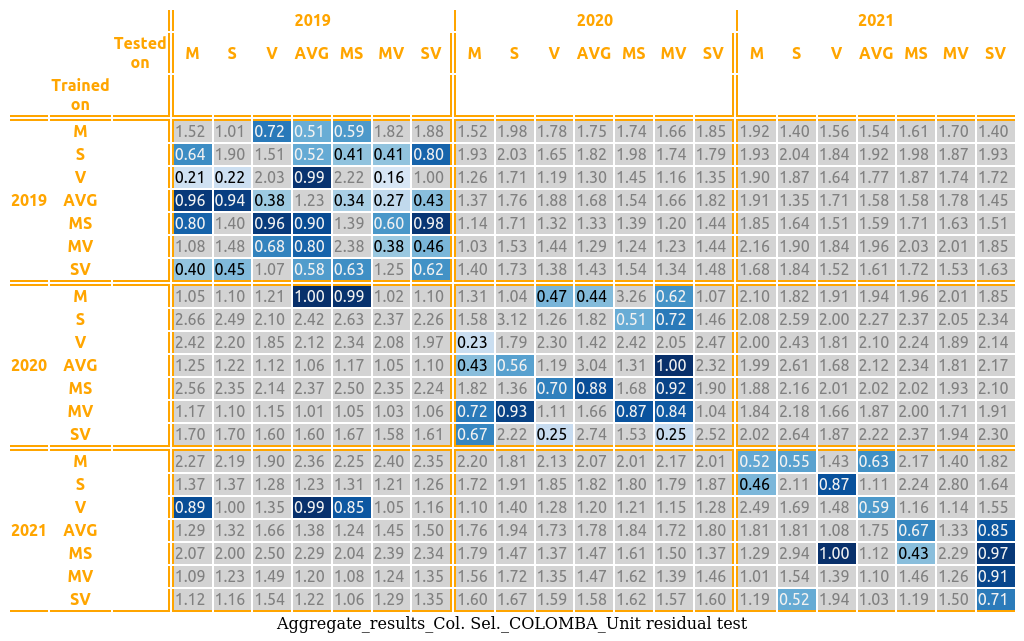}
            \includegraphics[width=0.98\linewidth, trim={0 0.7cm  0 0 },clip]{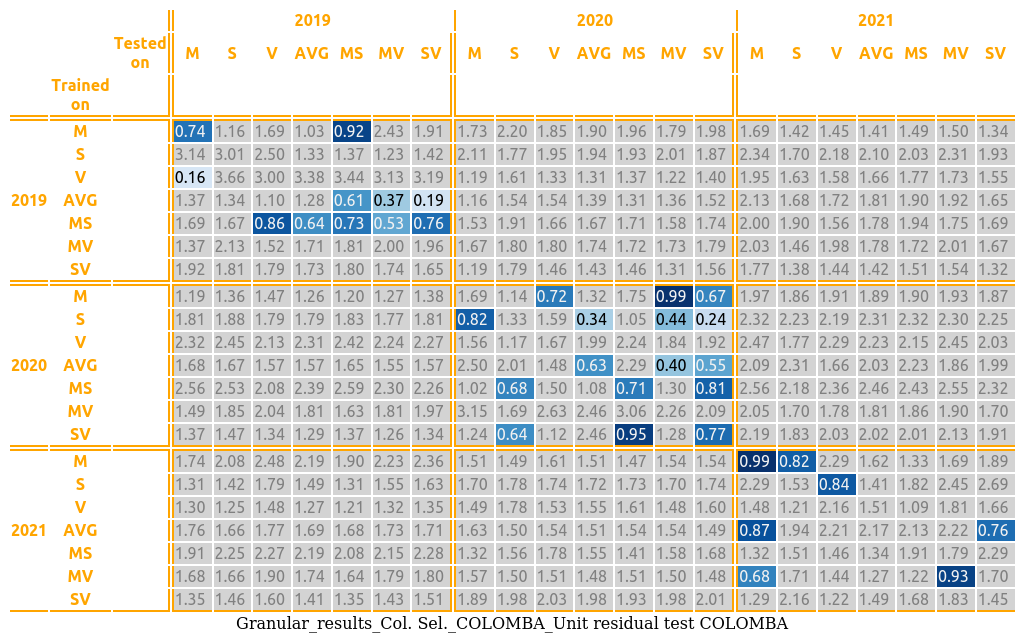  }
    \caption{Performance (SSE residual $r^{2}$) of the variety-specific Colomba model (top) and the variety-agnostic model tested on the Colomba variety (bottom). Rows correspond to the training dataset, columns show the testing configuration. SSE residuals $r^{2}\geq 1$ are grayed out.}
    \label{fig:COL_residuals_var_agn_var_spec}
\end{figure}

\begin{figure}
    \centering

        \includegraphics[width=0.98\linewidth, trim={0 0.7cm  0 0 },clip]{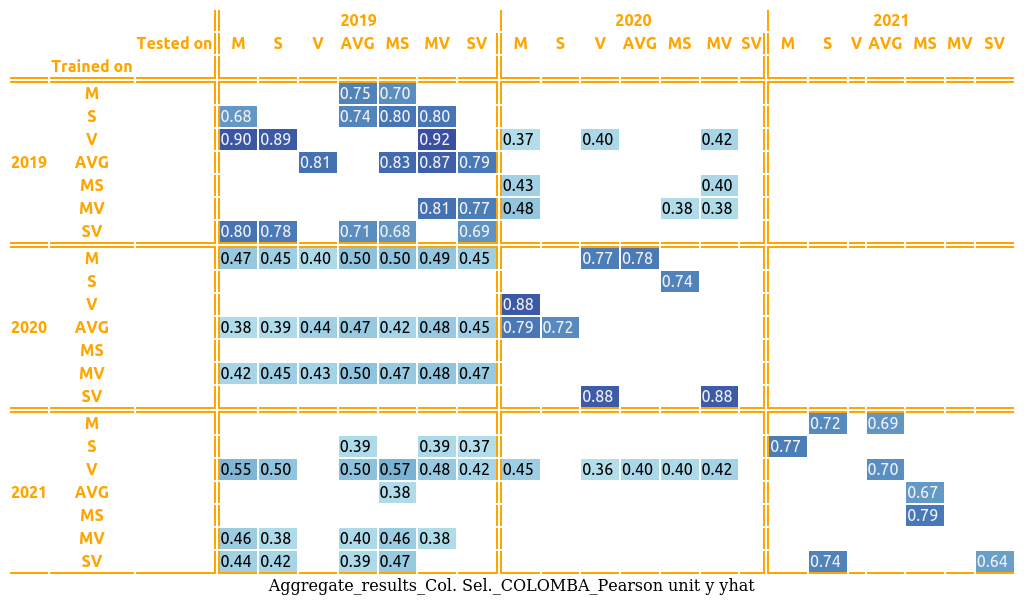}
            \includegraphics[width=0.98\linewidth, trim={0 0.7cm  0 0 },clip]{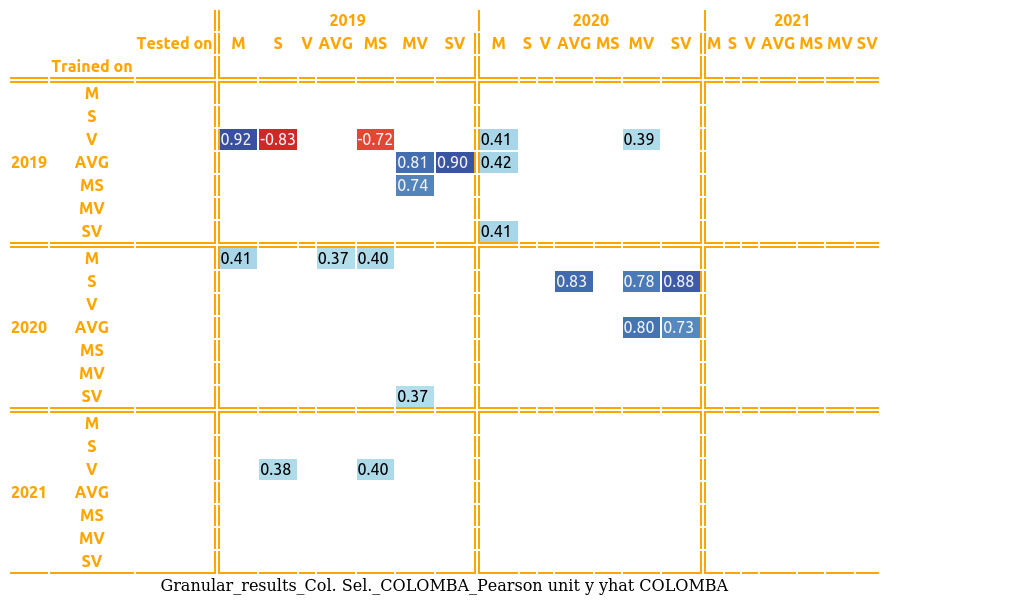}
    \caption{Performance (Pearson correlation $c$) of the variety-specific Colomba model (top) and the variety-agnostic model tested on the Colomba variety (bottom). Rows correspond to the training dataset, columns show the testing configuration. Correlations with $p\geq 0.05$ are omitted.}
    \label{fig:COL_pearson_corr_var_agn_var_spec}
\end{figure}

\begin{figure}
    \centering

        \includegraphics[width=0.98\linewidth, trim={0 0.7cm  0 0 },clip]{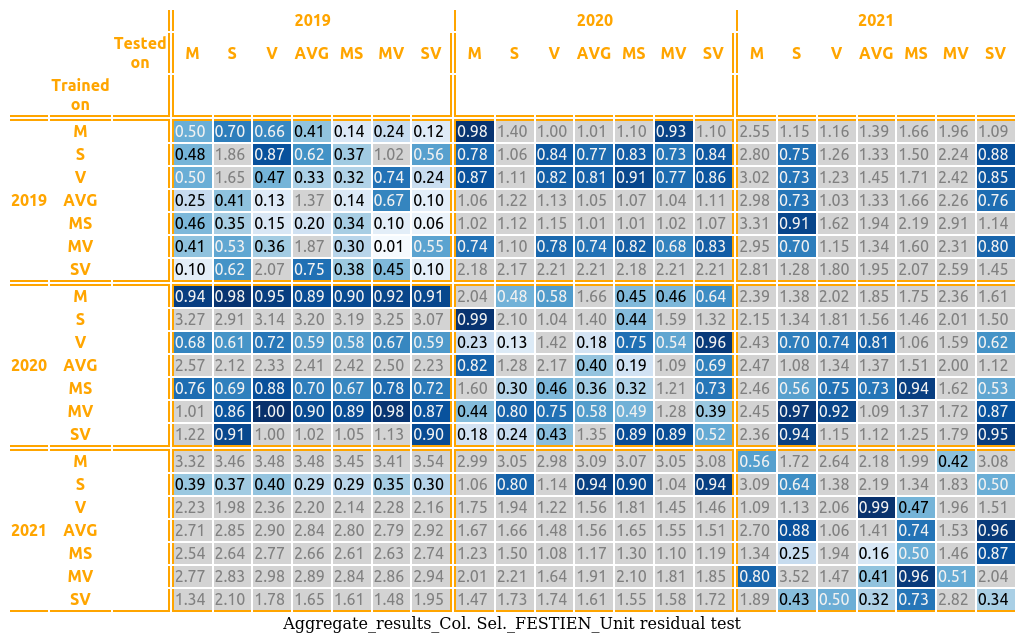}
            \includegraphics[width=0.98\linewidth, trim={0 0.7cm  0 0 },clip]{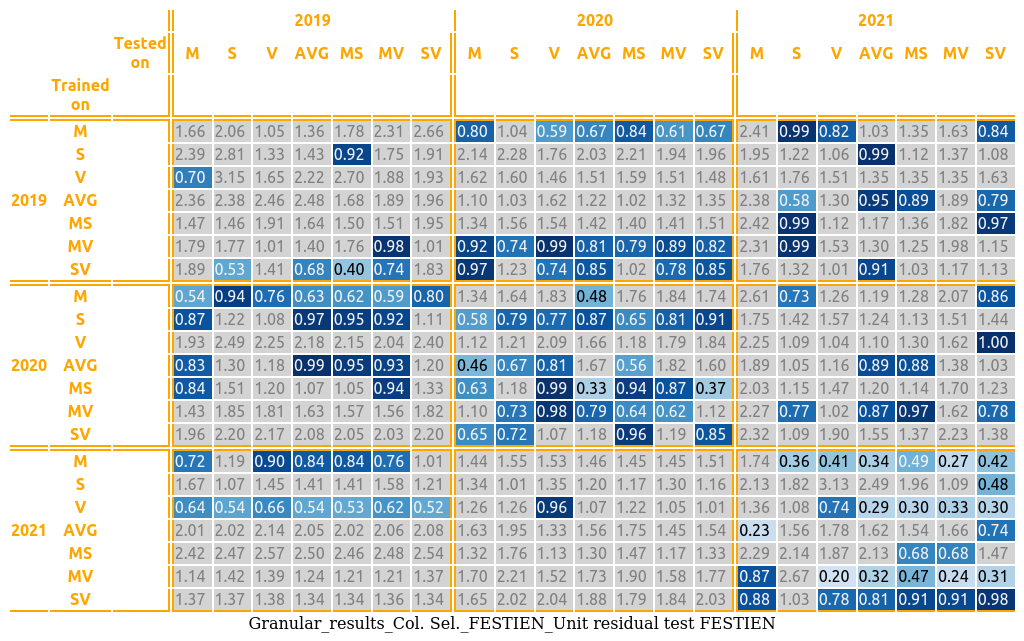  }
    \caption{Performance (SSE residual $r^{2}$) of the variety-specific Festien model (top) and the variety-agnostic model tested on the Festien variety (bottom). Rows correspond to the training dataset, columns show the testing configuration. SSE residuals $r^{2}\geq 1$ are grayed out.}
    \label{fig:FES_residuals_var_agn_var_spec}
\end{figure}

\begin{figure}
    \centering

        \includegraphics[width=0.98\linewidth, trim={0 0.7cm  0 0 },clip]{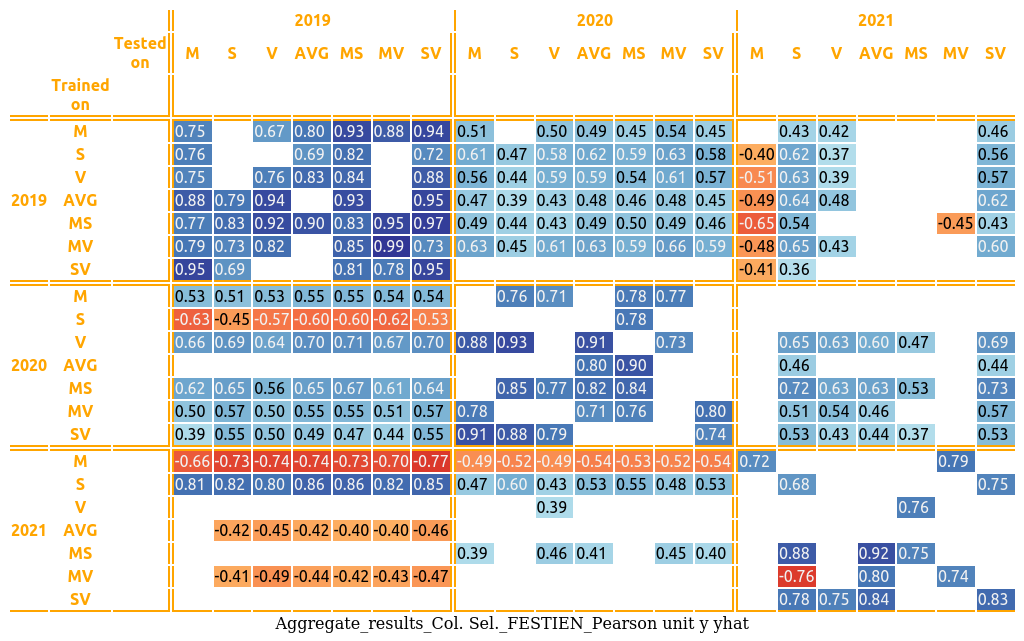}
            \includegraphics[width=0.98\linewidth, trim={0 0.7cm  0 0 },clip]{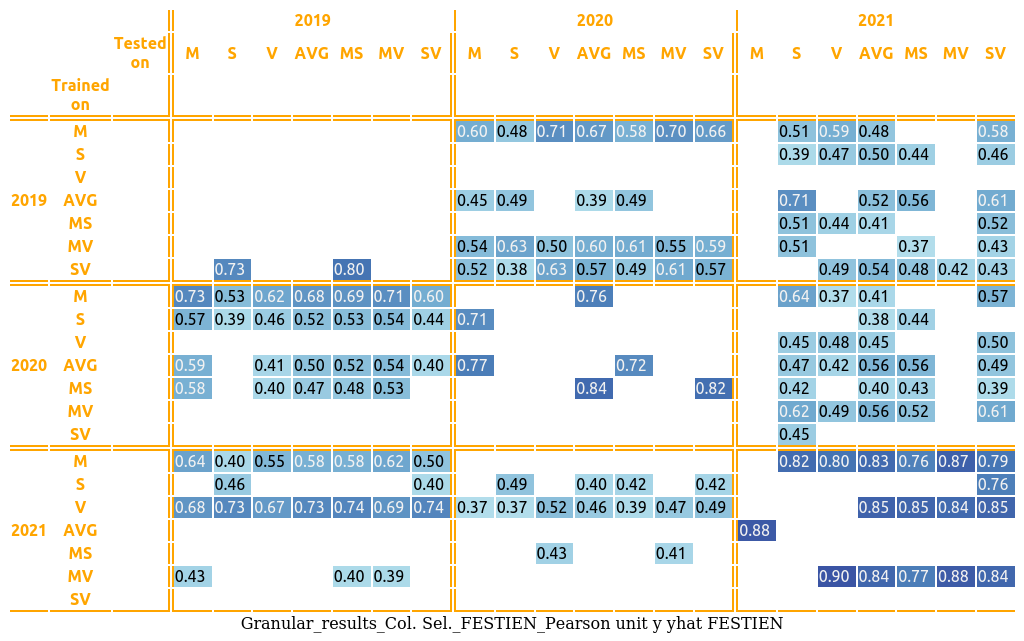}
    \caption{Performance (Pearson correlation $c$) of the variety-specific Festien model (top) and the variety-agnostic model tested on the Festien variety (bottom). Rows correspond to the training dataset, columns show the testing configuration. Correlations with $p\geq 0.05$ are omitted.}
    \label{fig:FES_pearson_corr_var_agn_var_spec}
\end{figure}

\begin{figure}
    \centering

        \includegraphics[width=0.98\linewidth, trim={0 0.7cm  0 0 },clip]{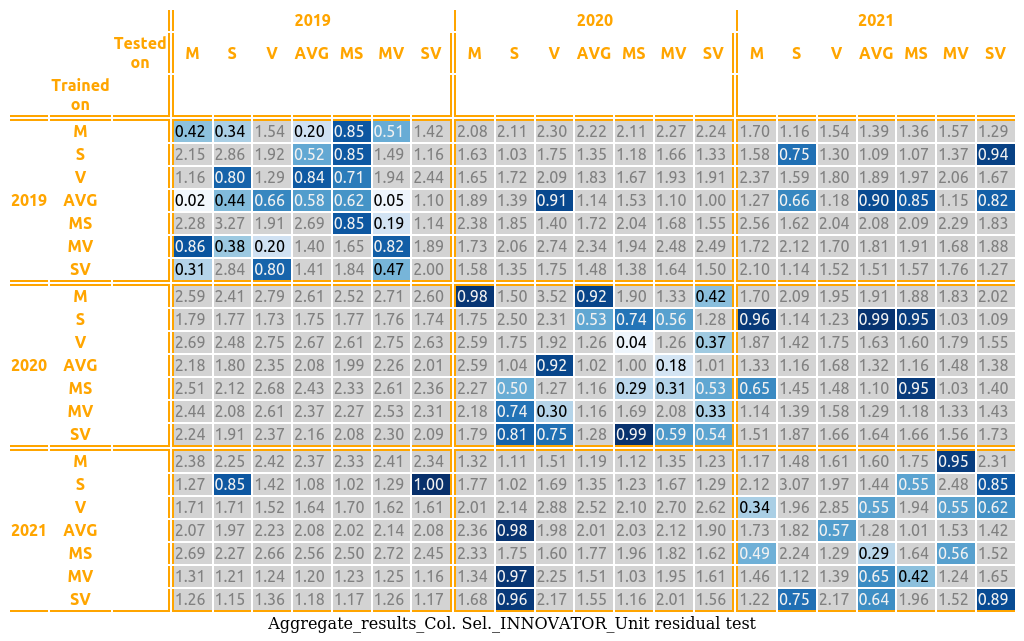}
            \includegraphics[width=0.98\linewidth, trim={0 0.7cm  0 0 },clip]{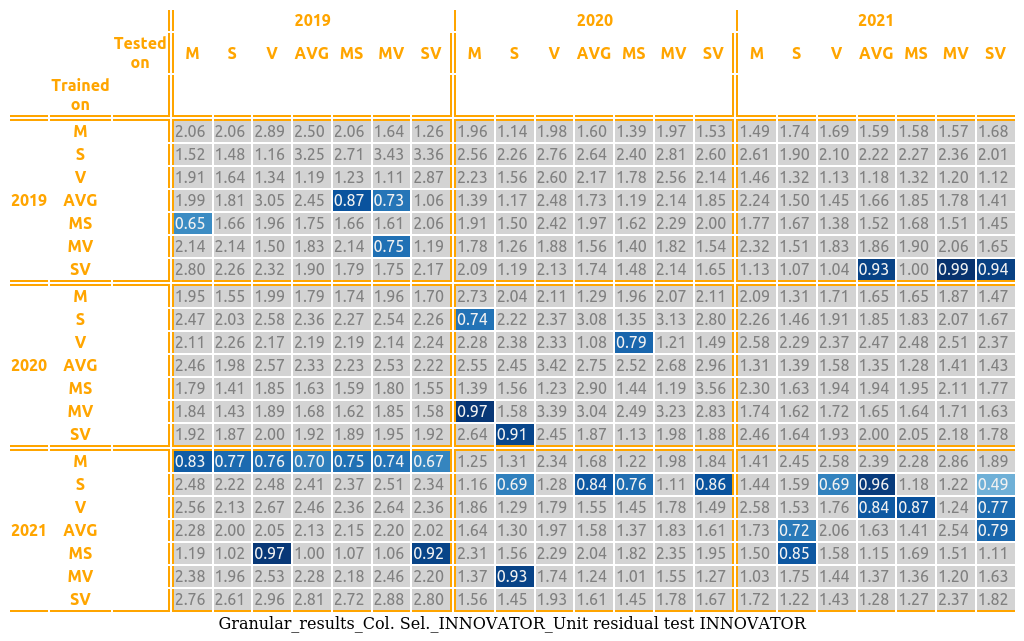 }
    \caption{Performance (SSE residual $r^{2}$) of the variety-specific Innovator model (top) and the variety-agnostic model tested on the Innovator variety (bottom). Rows correspond to the training dataset, columns show the testing configuration. SSE residuals $r^{2}\geq 1$ are grayed out.}
    \label{fig:INN_residuals_var_agn_var_spec}
\end{figure}

\begin{figure}
    \centering

        \includegraphics[width=0.98\linewidth, trim={0 0.7cm  0 0 },clip]{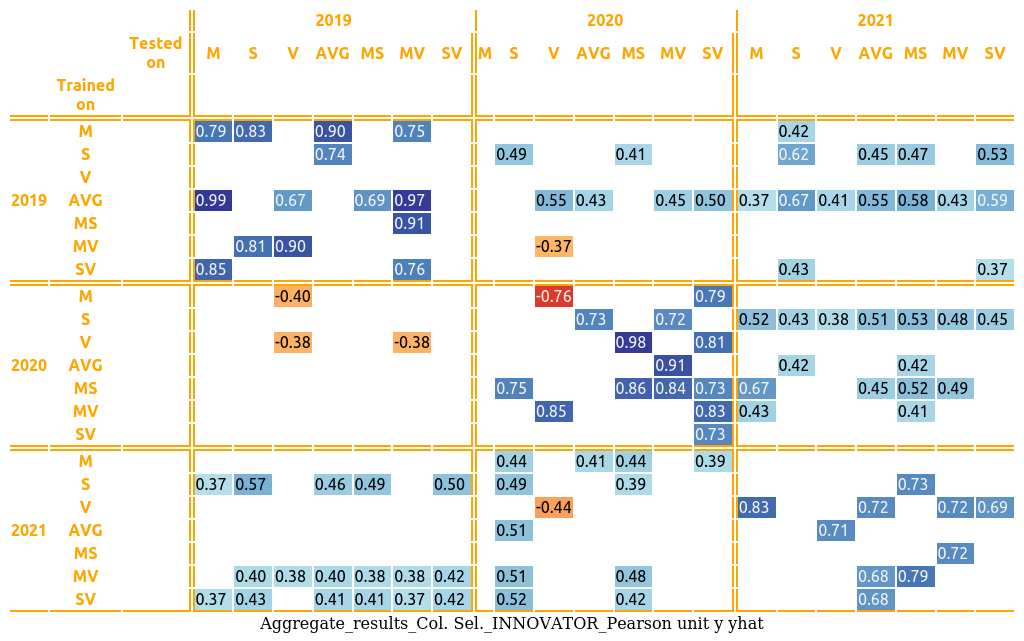}
            \includegraphics[width=0.98\linewidth, trim={0 0.7cm  0 0 },clip]{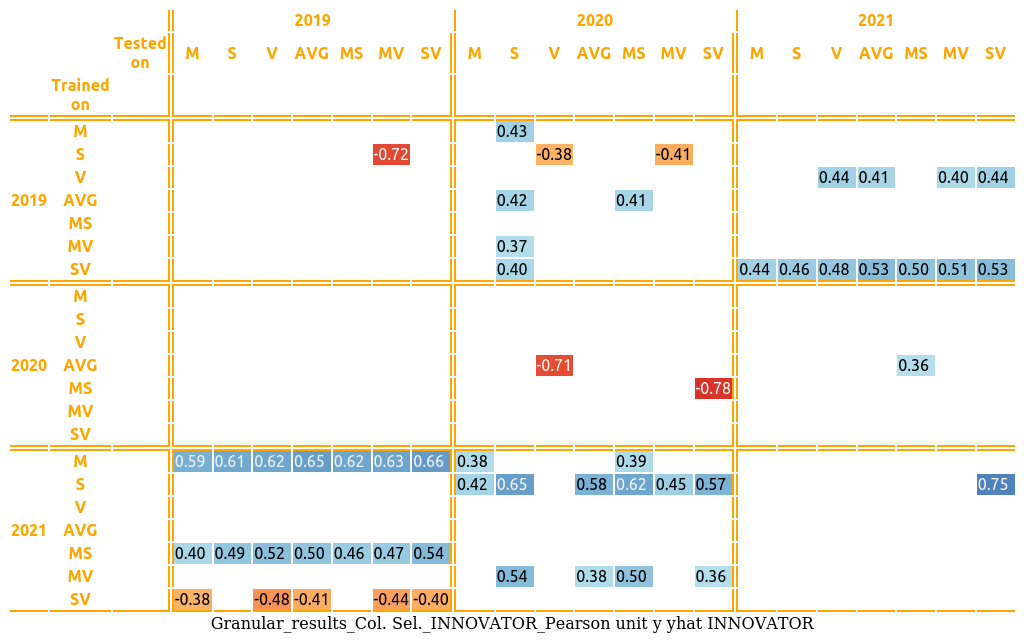}
    \caption{Performance (Pearson correlation $c$) of the variety-specific Innovator model (top) and the variety-agnostic model tested on the Innovator variety (bottom). Rows correspond to the training dataset, columns show the testing configuration. Correlations with $p\geq 0.05$ are omitted.}
    \label{fig:INN_pearson_corr_var_agn_var_spec}
\end{figure}

\begin{figure}
    \centering

        \includegraphics[width=0.98\linewidth, trim={0 0.7cm  0 0 },clip]{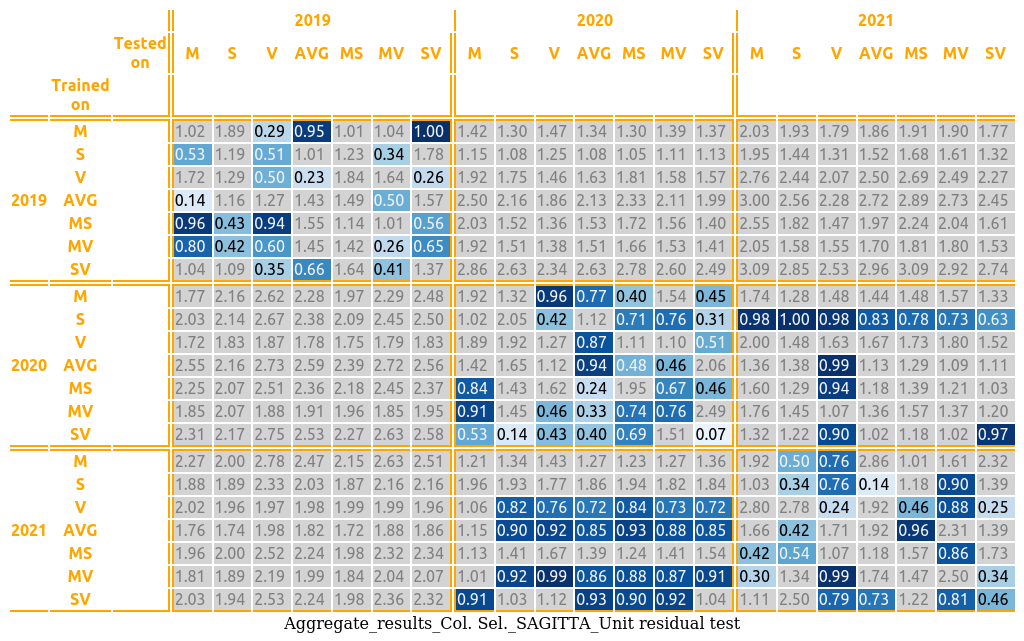}
            \includegraphics[width=0.98\linewidth, trim={0 0.7cm  0 0 },clip]{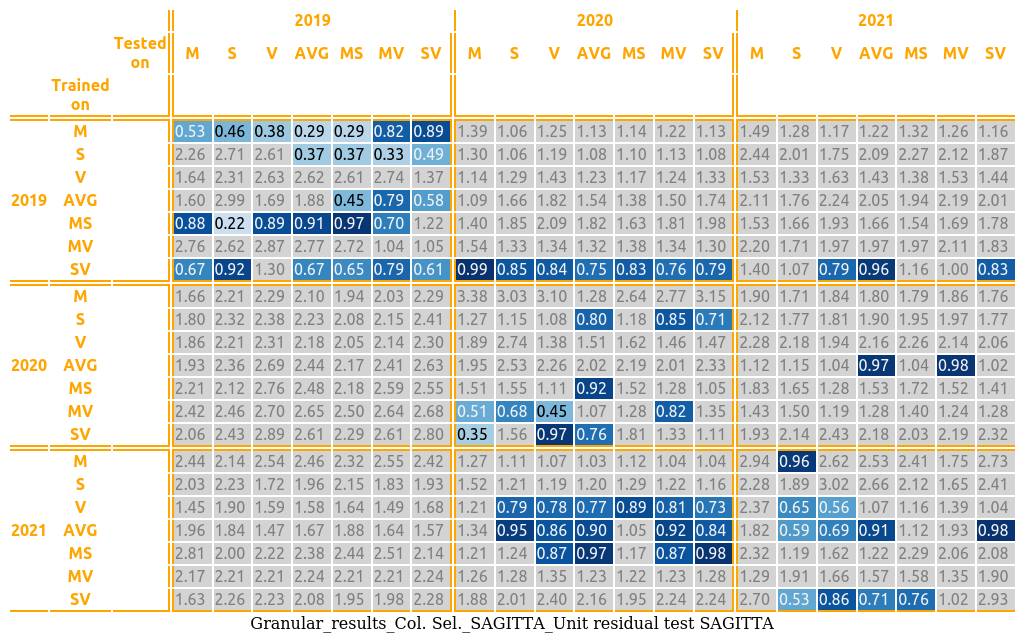 }
    \caption{Performance (SSE residual $r^{2}$) of the variety-specific Sagitta model (top) and the variety-agnostic model tested on the Sagitta variety (bottom). Rows correspond to the training dataset, columns show the testing configuration. SSE residuals $r^{2}\geq 1$ are grayed out.}
    \label{fig:SAG_residuals_var_agn_var_spec}
\end{figure}

\begin{figure}
    \centering

        \includegraphics[width=0.98\linewidth, trim={0 0.7cm  0 0 },clip]{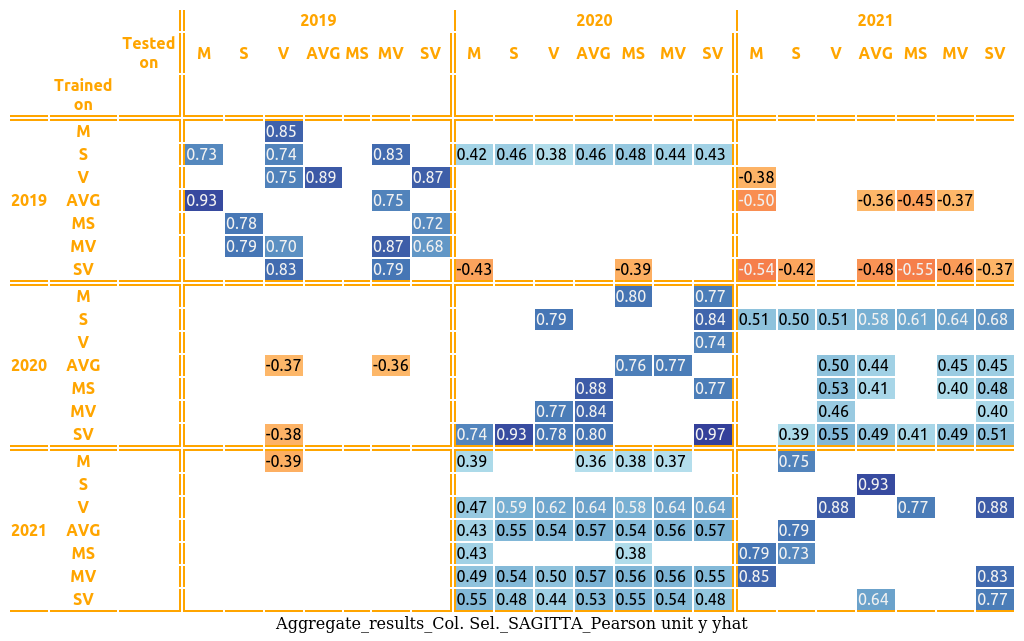}
            \includegraphics[width=0.98\linewidth, trim={0 0.7cm  0 0 },clip]{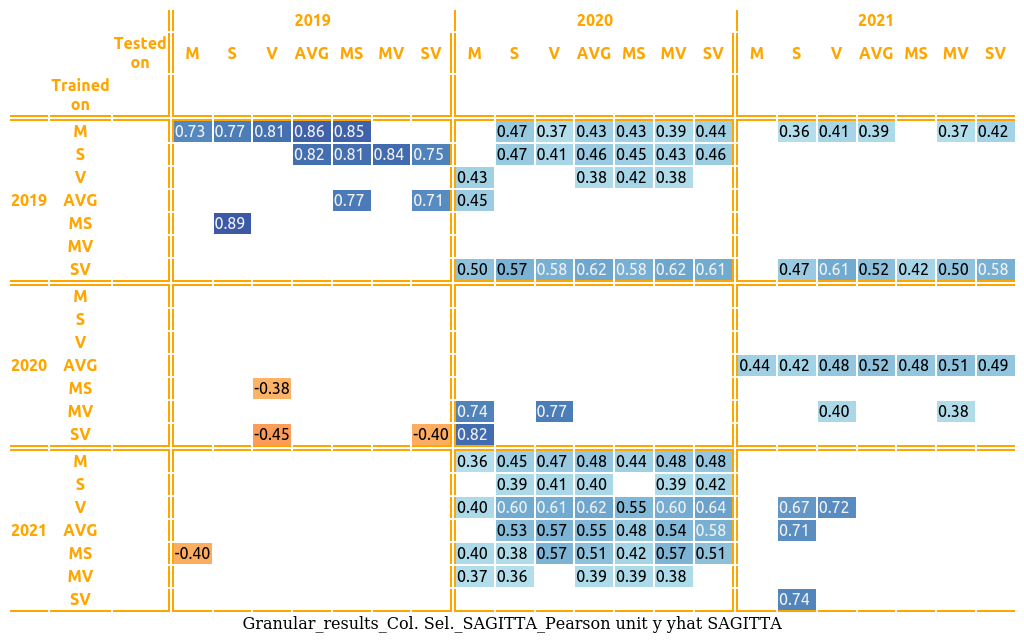}
    \caption{Performance (Pearson correlation $c$) of the variety-specific Sagitta model (top) and the variety-agnostic model tested on the Sagitta variety (bottom). Rows correspond to the training dataset, columns show the testing configuration. Correlations with $p\geq 0.05$ are omitted.}
    \label{fig:SAG_pearson_corr_var_agn_var_spec}
\end{figure}

\begin{figure}
    \centering

        \includegraphics[width=0.98\linewidth, trim={0 0.7cm  0 0 },clip]{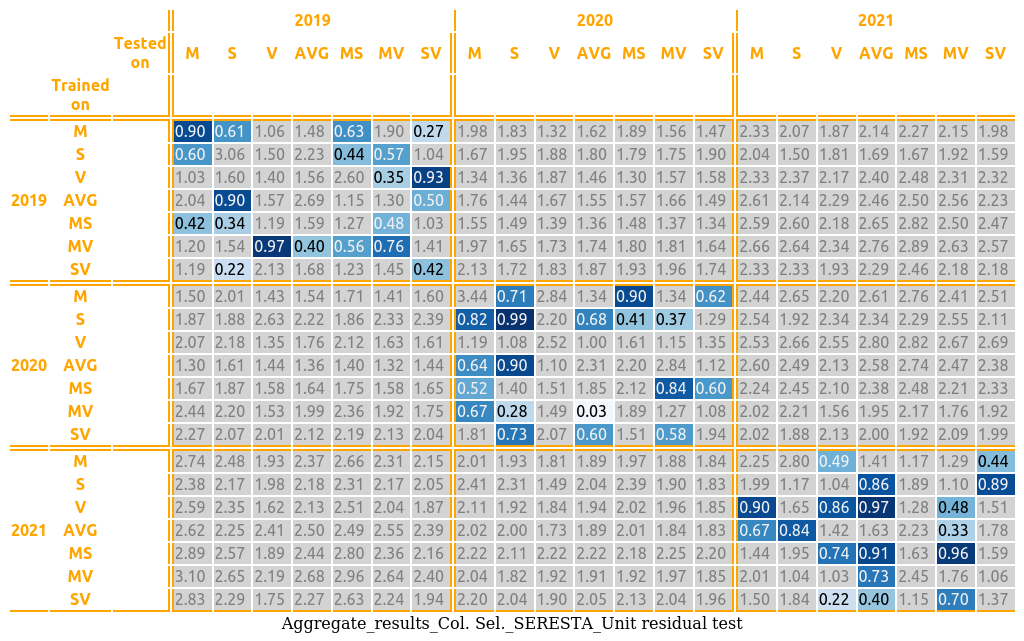}
            \includegraphics[width=0.98\linewidth, trim={0 0.7cm  0 0 },clip]{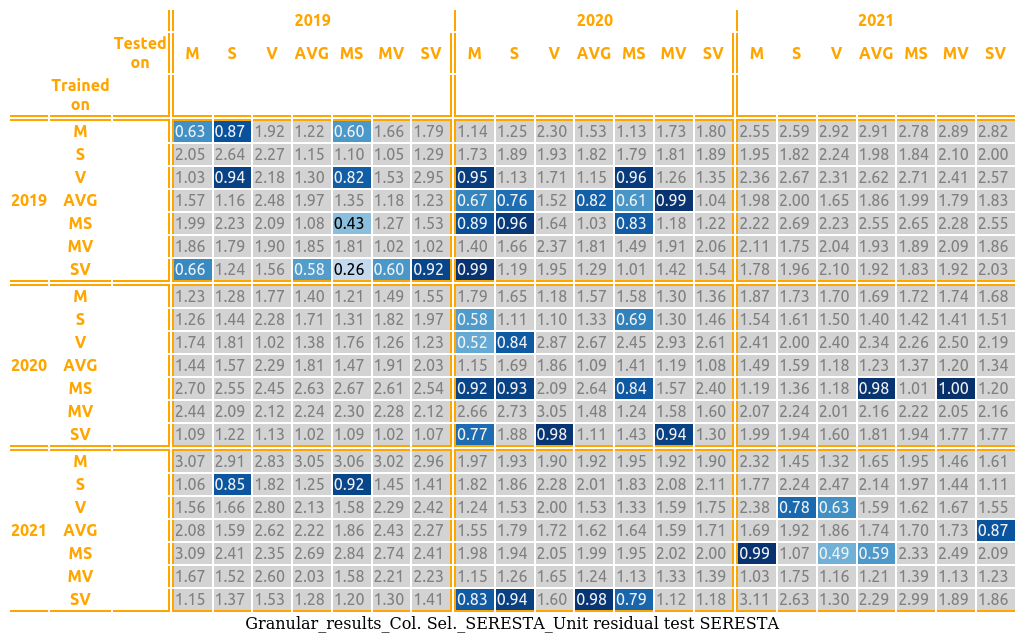}
    \caption{Performance (SSE residual $r^{2}$) of the variety-specific Seresta model (top) and the variety-agnostic model tested on the Seresta variety (bottom). Rows correspond to the training dataset, columns show the testing configuration. SSE residuals $r^{2}\geq 1$ are grayed out.}
    \label{fig:SER_residuals_var_agn_var_spec}
\end{figure}

\begin{figure}
    \centering

        \includegraphics[width=0.98\linewidth, trim={0 0.7cm  0 0 },clip]{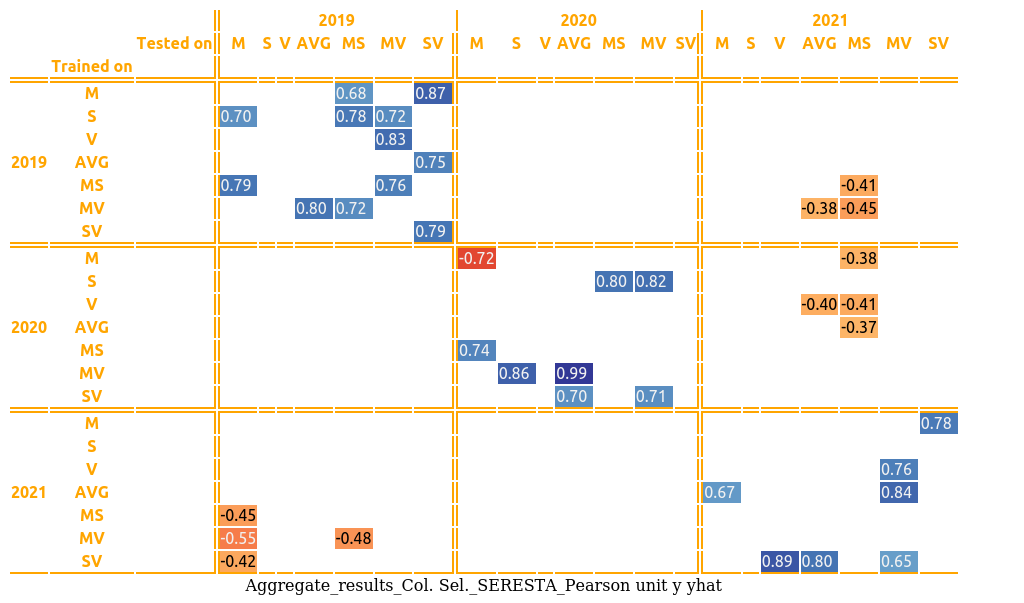}
            \includegraphics[width=0.98\linewidth, trim={0 0.7cm  0 0 },clip]{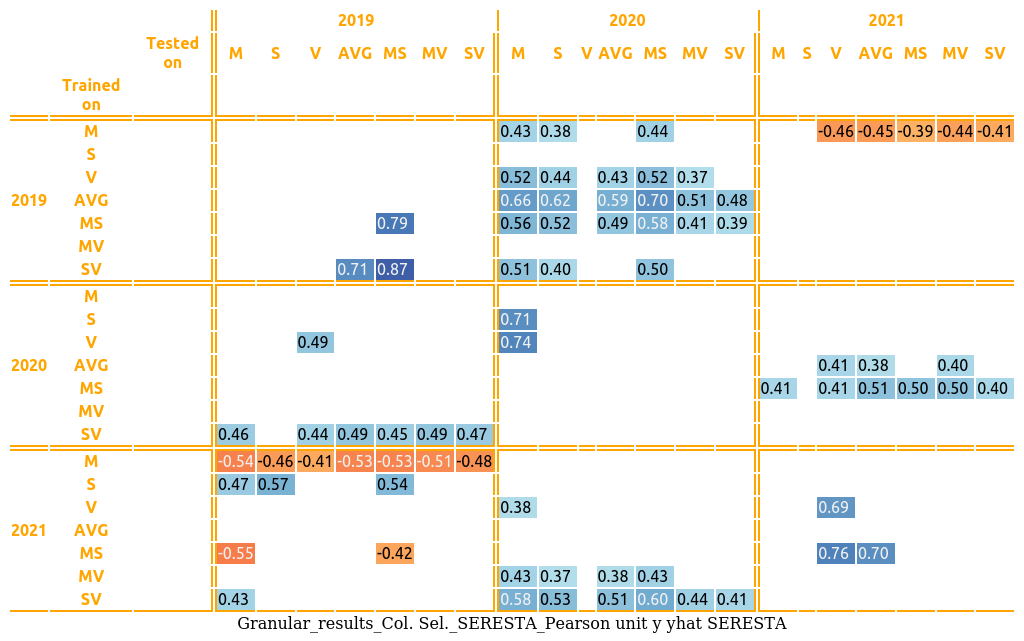}
    \caption{Performance (Pearson correlation $c$) of the variety-specific Seresta model (top) and the variety-agnostic model tested on the Seresta variety (bottom). Rows correspond to the training dataset, columns show the testing configuration. Correlations with $p\geq 0.05$ are omitted.}
    \label{fig:SER_pearson_corr_var_agn_var_spec}
\end{figure}

Since in practice a single variety is usually planted in the field by potato farmers, it is important to know how well one can predict the vigor of a seedlot not across all varieties but within its specific variety. The assumption behind the variety-agnostic model is that the biochemical mechanisms behind a weak or a strong vigor are the same for all potato varieties. However, it is also possible that these mechanisms depend on the variety. Therefore, six separate variety-specific IR models have also been trained and the results are presented in Figures~\ref{fig:CHA_residuals_var_agn_var_spec}--\ref{fig:SER_pearson_corr_var_agn_var_spec} next to the corresponding segregated results of the variety-agnostic model. 

From Figures~\ref{fig:CHA_residuals_var_agn_var_spec}--\ref{fig:SER_pearson_corr_var_agn_var_spec}, it is clear that the quality of predictions within the varieties is much lower than across the whole set of varieties, Figure~\ref{fig:var_agn_full_field}, which confirms the strong genotype effect on the vigor of seedlots.  
Apart from the statistically less reliable within-the-year results (diagonal blocks in the tables), for the four of the six varieties (Challenger, Innovator, Seresta, Colomba) neither the variety-agnostic nor the variety-specific models are able to explain the intra-varietal variations in the vigor when presented with the tuber data of a different year. Although, the correlations between the predicted and the measured vitalities do fall just a little short of 50\% for these varieties, this corresponds to the residuals with the magnitude above one, i.e., such predictions are practically useless.

The notable exception is the Festien variety, which exhibits a fairly predictable behavior, see Figures~\ref{fig:FES_residuals_var_agn_var_spec}, \ref{fig:FES_pearson_corr_var_agn_var_spec}, with the variety-specific model showing somewhat better performance if compared to the variety-agnostic model. The variety-specific Festien model trained on the tuber data of the year 2020 and the average vigor data of the M and S fields in 2020 predicts/explains 30\% of the variation of the three-field average vigor in the year 2019 and 27\% of the three-field average vigor in 2021, when presented with the corresponding seed tuber data.

In the second place is the Sagitta variety, which shows a similar partially predictable behavior between the years 2020 and 2021, see Figures~\ref{fig:SAG_residuals_var_agn_var_spec}--\ref{fig:SAG_pearson_corr_var_agn_var_spec}. However, this predictability does not extend to the year 2019 for the Sagitta variety.

\subsection*{Predictive power of different tuber data types}

The IR model provides a measure of the usefulness of each particular predictor variable \cite{Atza-Budko-2024}. 
When a new predictor-variable dataset (tuber properties) is considered for prediction purposes, the IR method can perform a `prediction' of this known dataset by the model created on the training dataset of tuber properties. For example, the dataset of tuber properties in the year 2021 can be predicted from the dataset of tuber properties in the year 2020. The quality of the $X$-data prediction is measured by the normalized residuals for each individual predictor variable, i.e., abundance of a metabolite, spectral amplitude of the FTIR or HSI signature, abundance of an inorganic element. The smallness of such residuals is the necessary condition for the usefulness of the particular predictor variable. One can only achieve a good prediction of the dependent variable (vigor) if these residuals are small.

\begin{figure}
    \centering
    \includegraphics[width=0.99\linewidth,trim={4.5cm, 1.5cm 5cm 4cm}, clip]{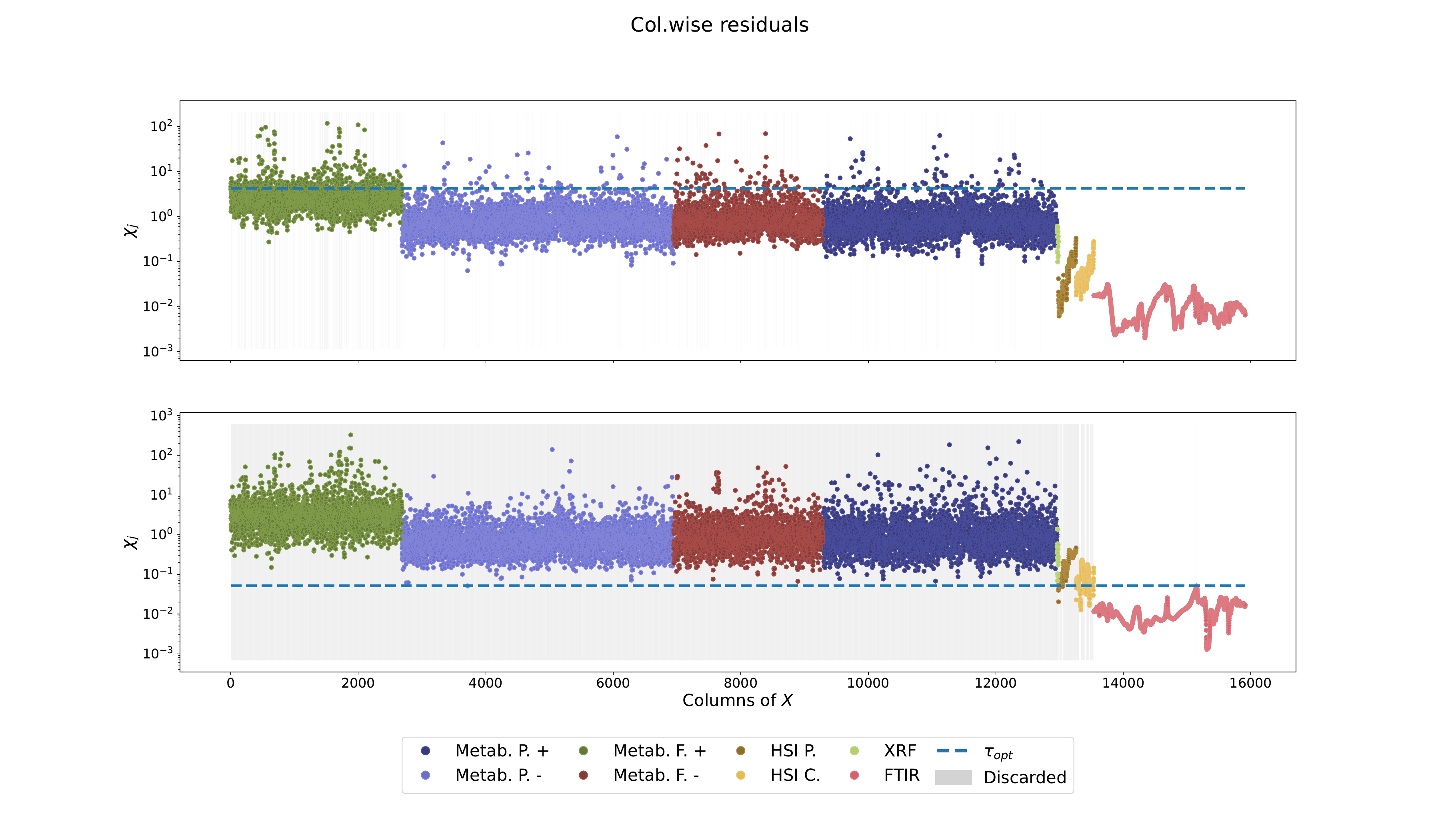}
    \caption{The $X$-data residuals for the variety-agnostic (top) and variety-specific (bottom) 2020 models tested on the 2021 Festien tuber data. Colors indicate the data type (see legend). Dashed horizontal lines show the feature selection thresholds, with the rejected predictors marked by vertical gray lines.}
    \label{fig:col-wise-residuals}
\end{figure}

The predictor-variable residuals, or the $X$-data residuals, can be used in two ways: to rank the predictive power of the different data types, and to perform feature selection. Figure~\ref{fig:col-wise-residuals} shows the $X$-data residuals grouped and colored by the type of the tuber data for the training dataset of the year 2020 and the testing dataset of the year 2021. In the top plot, the residuals of the 2020 variety-agnostic model predicting the 2021 Festien $X$-data are shown, and in the bottom plot, the residuals for the prediction of the same $X$-data by the 2020 variety-specific Festien model are displayed. In this and other training-testing configurations, the FTIR dataset consistently shows the smallest residuals, followed by the two HSI datasets, the XRF and the metabolome datasets. 

\begin{figure}
    \centering
    \includegraphics[width=0.98\linewidth, trim={4cm 3cm 0cm 4cm}, clip]{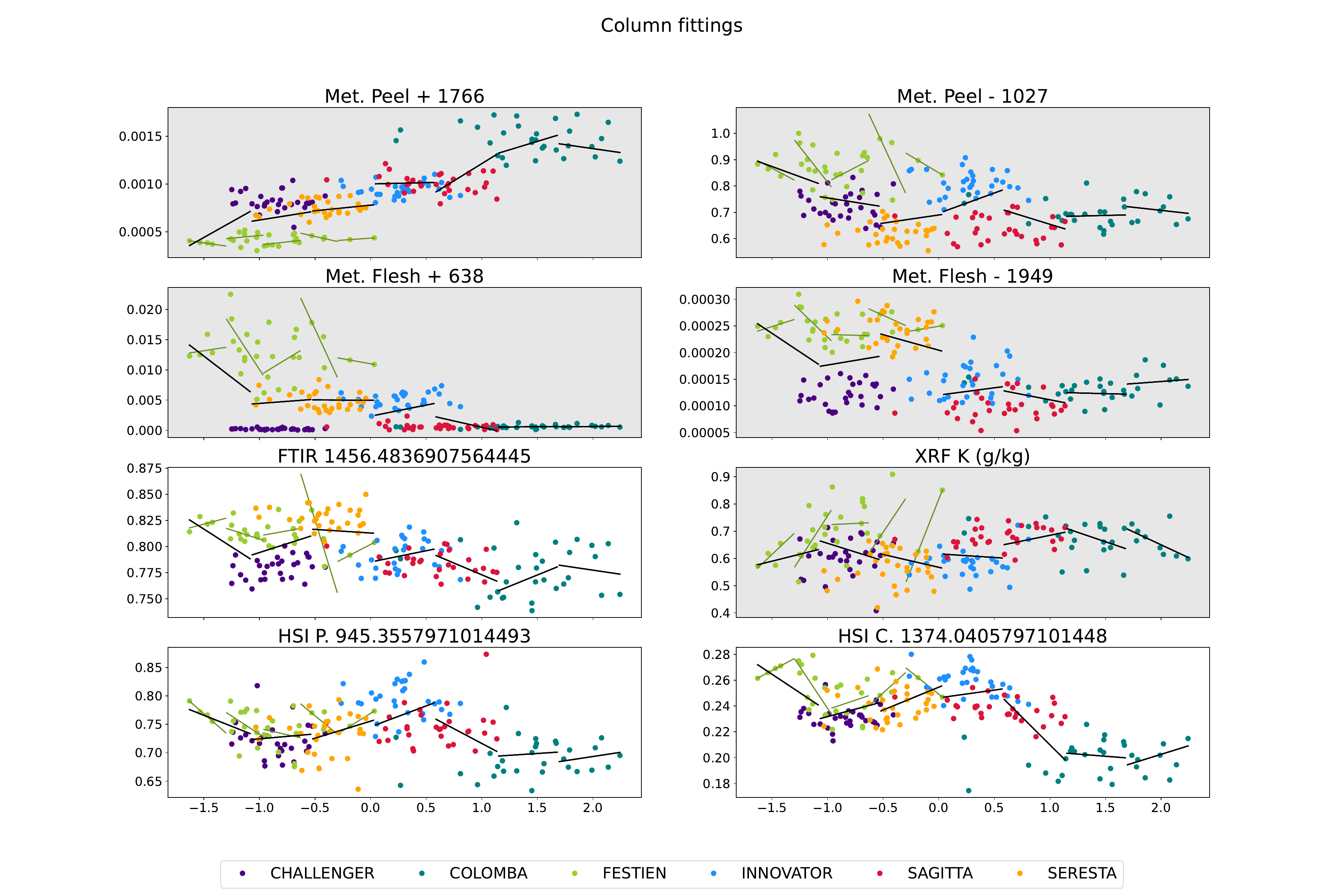}
    \caption{Individual predictor data (vertical axes) fitted by piece-wise linear functions of the vigor parameter (horizontal axis). For each data type we show the predictor with the lowest residual in the Festien specific model. The training was performed on the average vigor of the three test fields and the resulting fits by the variety-agnostic model are shown as broken solid black lines. The fits produced by the variety-specific Festien model are shown as broken light-green lines. The data type is indicated in the titles of the plots together with the predictor label. Plots with the light-gray background show predictors that have been rejected by the variety-specific Festien model.}
    \label{fig:best_col_fits}
\end{figure}
The metabolome data, on average, demonstrate higher residuals, with some metabolites producing smaller residuals that are still higher than the residuals of the FTIR data. In mathematical terms, this means that the training metabolome dataset is less complete than the training FTIR dataset \cite{Atza-Budko-2024}. There can be several reasons why the IR algorithm indicates such incompleteness. First, the dependence of the majority of metabolites on the vigor parameter is of `non-functional' type, i.e., does not look like a sampled graph of a function. This is indeed what is observed in the trained variety-agnostic IR model. In Figure~\ref{fig:best_col_fits} one can see the fits produced by the IR algorithm for the various predictor variables. The horizontal axis is the vigor parameter, and the vertical axis is the value of the corresponding predictor. The data points are colored in accordance with the seed tuber variety. The metabolome predictors shown in Figure~\ref{fig:best_col_fits} (top four plots) exhibits a typical nonfunctional behavior \cite{Atza-Budko-2024}, since the data points from different varieties represent vertically shifted clusters, i.e., it is impossible to draw a graph of a mathematical function of the vigor variable through these clusters.

The second reason for the apparent incompleteness of the metabolome dataset, that can observed in the variety-specific Festien model, is the relatively high noise in the data, where the data points appear to form a cloud rather than being aligned along a graph of a function, see Figure~\ref{fig:best_col_fits} (left column, second plot from the top, green point cloud). The noise in these plots comes from two sources: the vigor data (error in the point location along the horizontal axis) and the metabolome data (error in the point location along the vertical axis). It is not clear which of the two noise sources is dominant, however, it is clear that the noise in the metabolome data is higher than in, e.g., the FTIR or HSI data.

\section*{Discussion}

In this project, it has been confirmed that the potato plants grown from genetically identical seed tubers of different production origin and physiological state show systematic differences in vigor. These differences have been quantified here as significant correlations in the canopy area of seedlots across test fields. The answer to the question about the possibility to predict the seedlot vigor from the biochemical properties of the seed tubers is ``{\it sometimes yes}''. The vigor of two varieties, Festien and to a lesser extent Sagitta, appears to be fairly predictable. Although, the model for the Sagitta variety fails to predict the vigor in one of the three test years. Notably, it was also the year which showed relatively small correlations in vigor between the test fields for this variety. Festien and Sagitta were also the better predictable varieties in our previous microbiome study \cite{Yang-et-al-2024}, where it was shown that this level of predictability is enough to distinguish between the three practically meaningful classes for the seedlots of these varieties: below average, average, and above average. Moreover, the feature selection algorithm applied to all biochemical tuber data, indicates that it is sufficient to measure the FTIR spectra of the dry samples of the seed tubers to achieve this classification.

For the other four varieties (Challenger, Colomba, Innovator, and Seresta) the vigor of the seedlots could not be predicted in any reliable way. It is worth noting that the vigor correlations across the test fields were also not as consistent for these varieties, which may be a sign of a purely stochastic nature of vigor variations within these varieties. Nevertheless, in our opinion, the reasons behind this lack of predictability are not entirely clear. On the one hand, it is possible that the biochemical composition of the seed tubers simply does not define the vigor of these varieties to the same extent as with the Festien variety, and there are other processes at play that were not measured in the present experiment. A seed tuber is a slowly changing dynamical system. Therefore, measuring its biochemical constitution at a few points in time may deliver better predictors of the potato plant vigor. This could mean that the tubers of Festien are somehow more `stationary' than the tubers of other varieties.

On the other hand, these conclusions are also subject to the choice of the vigor parameter -- leaf canopy area on a selected date. Since the selected date was at the end of the exponential canopy growth period, even small variations in the emergence date or the rate of growth could lead to significant changes in the ordering of the seedlots at the measurement date, thus, reducing the correlation in the vigor parameter between the fields and the eventual predictability of vigor. It is possible that a different, possibly dynamic, vigor measure, while requiring more phenotyping effort, could be both more stable and more predictable.

Finally, observations in the year 2021 indicate a strong influence of the environment on some seedlots and varieties, which can be taken as the evidence of the genotype-by-environment interaction. The definition of the potato plant vigor and the quality of its prediction can be significantly improved if the influence of the environment is taken into account by incorporating the readily available environmental data into a physiology-informed growth model. This, however, requires a separate set of dedicated experiments or suitable historical data, in order to learn the genotype-specific environmental component and also calibrate the model for each particular genotype. One way forward is to incorporate the biochemical tuber data into the available physiology-informed models \cite{potato-models} calibrated for specific potato varieties.

\section*{Methods}

\subsection*{Measuring the plant vigor}
\subsubsection*{Drone-image segmentation and spatial correction}

The growth of the potato plants was monitored with the drone-mounted camera and the drone operator company delivered the stitched images (orthophotos) of each field taken approximately once per week. The leaf canopy of the potato plants was extracted from the images and the average canopy area per plant for each seedlot was calculated. The general image processing protocol and the example code demonstrating the extraction of the vigor data in the years 2019 and 2020 has been published in connection with the previous paper \cite{Yang-et-al-2024, micro-vigor-dataset}. The results presented here include the canopy data from the year 2021, the extraction of which required additional processing steps. For uniformity, the new processing pipeline was then retrospectively applied to the years 2019 and 2020 as well. The updated detailed protocol for the extraction of the canopy area and the corresponding vigor and seed tuber data are available online \cite{dataset_this_paper}.  

One of the image processing steps is the semi-automatic procedure that detects the boundaries of the seedlot plots. This was done in the image taken on the date when it was easy to detect the plots and distinguish between them. Due to the large scale of the experiment, the planting was performed from a moving tractor, which resulted in the plots having slightly irregular boundaries, that had to be represented by polygons rather than simple rectangles, see Figure~\ref{fig:plots_images}. Moreover, the irregularity of the planting scheme required some manual input while detecting the plot boundaries. To avoid repeating this laborious procedure in every snapshot of the same field, the field images were transformed to the same spatial reference frame with the help of permanent markers that were installed on the ground at the time of planting and later automatically detected in all images.

Once the plots were identified and assigned to the corresponding seedlots, the plot sub-images, see Figure~\ref{fig:plots_images}, were segmented by color to identify and count the plant pixels. This is achieved on the HSV representation of the image by selecting pixels for which the weighted average of some vegetation indices exceeds a threshold. The weights and thresholds are provided per measurement date in the protocol \cite{dataset_this_paper}. The raw canopy area for each plot is found by counting the identified canopy pixels and converting the result to $\text{cm}^2$ with the help of the known inter-ridge distances in each field. In the final step, the spatial trend induced by the test-field heterogeneity was removed from the raw canopy area data with the specialized R-package SpATS \cite{spats}, resulting in 180 canopy surface estimates per test field snapshot.

\subsubsection*{Choice of the vigor parameter}

From the observational point of view, the relative difference in the vigor of potato plants can manifest itself in several ways. First, some seed tubers can sprout sooner than others. Second, the sprouts can grow at  different rates, emerging above the ground at different times. Finally, the rates of growth of the leaf canopies can be different, so that canopies achieve different sizes at the moment when the plant blooms and produces the tubers. While the first two processes are certainly indicative of the seed tuber vigor, the size of the canopy at the time of tuber bulking is more important for the final tuber yield of the potato plant. That is why, in this project, we focused on the canopy size at a fixed date, later in the season.

\begin{figure}
    \centering
        \includegraphics[width=0.98\linewidth, trim={3px, 0.67em, 3px, 3cm}, clip]{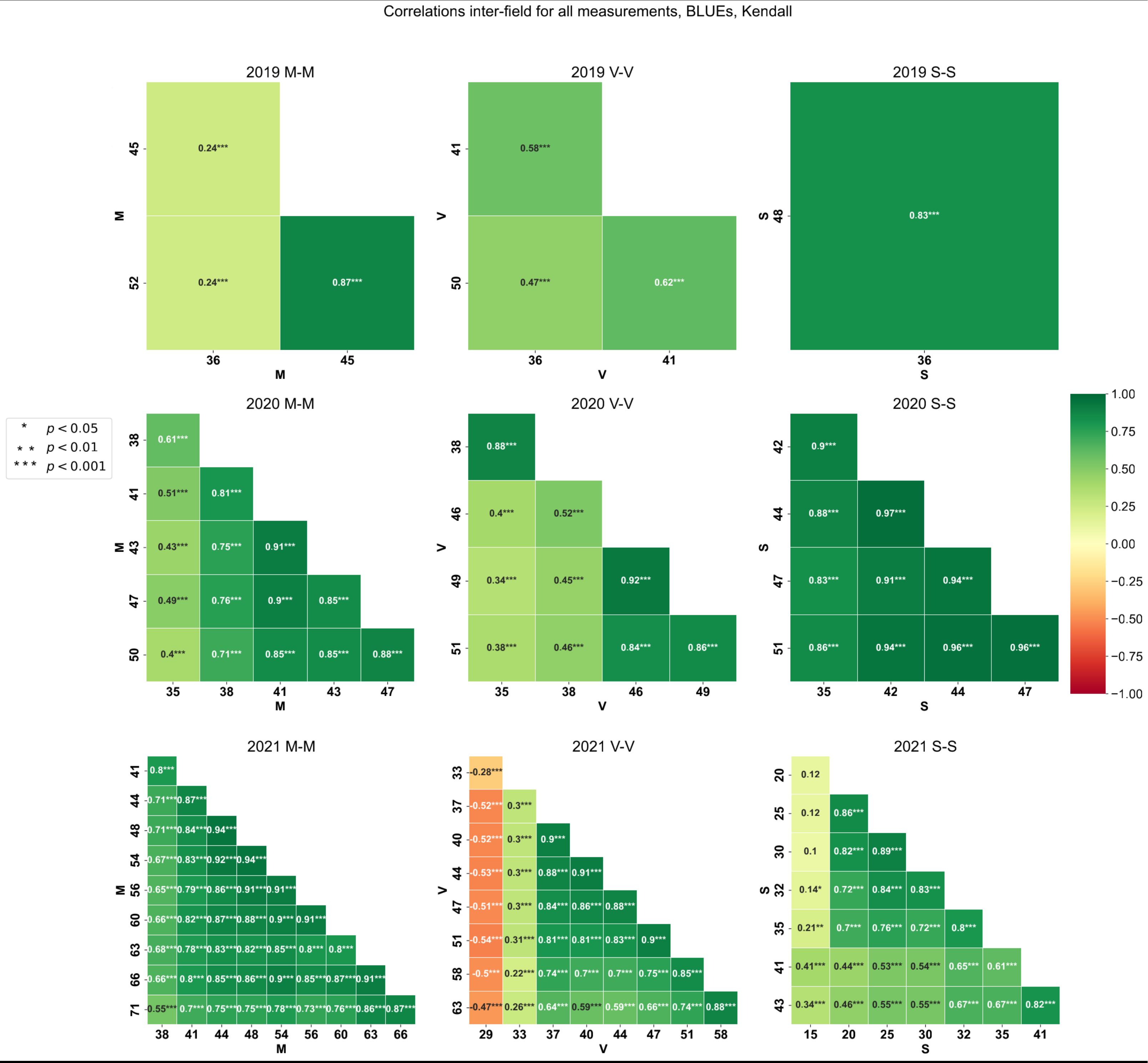}
    \caption{Kendall's $\tau$ coefficient of the canopy sizes between the measurement dates (DAPs) in each test field and year. In this case, the Kendall $\tau$ coefficient measures the stability of the seedlot order over three quantiles (low, middle, high) and would be equal one, if the seedlot canopies did not change their quantiles between the dates.}
    \label{fig:all_k_corr}
\end{figure}

By analyzing the canopy images taken on consecutive dates, it was found that the seedlots achieve a stable order
in terms of the size of their canopy size towards the end of the exponential growth period, see Figure~\ref{fig:all_k_corr}. The date which showed a large (Kendall) correlation with the previous dates and, simultaneously, a significant variance between the canopies of the variety, was chosen as the date when the canopy area was taken to be the vigor parameter of the seedlot. The second condition is necessary because eventually the canopies of the neighboring plants merge, become indistinguishable in the drone images, and start competing for the incoming light.

It appears that in most cases the 47-th Day After Planting (DAP) is the best choice, since the plants of all six varieties have completely emerged, the canopy-based ordering of the seedlots has stabilized, and the variations between the seedlots are still significant. In the year 2019, with only a few dates available, the choice was mostly dictated by the seedlot order stability. In 2020 and 2021 the 47-th DAP was chosen where available and in the test field V in 2020, the choice was again based on the date of maximal order stability. The full list of selected dates is provided in the Table~\ref{tab:choice_dap}.

\begin{figure}
    \centering
    \includegraphics[width=0.98\linewidth, trim={3px, 0.1em, 3px, 5.4cm}, clip]{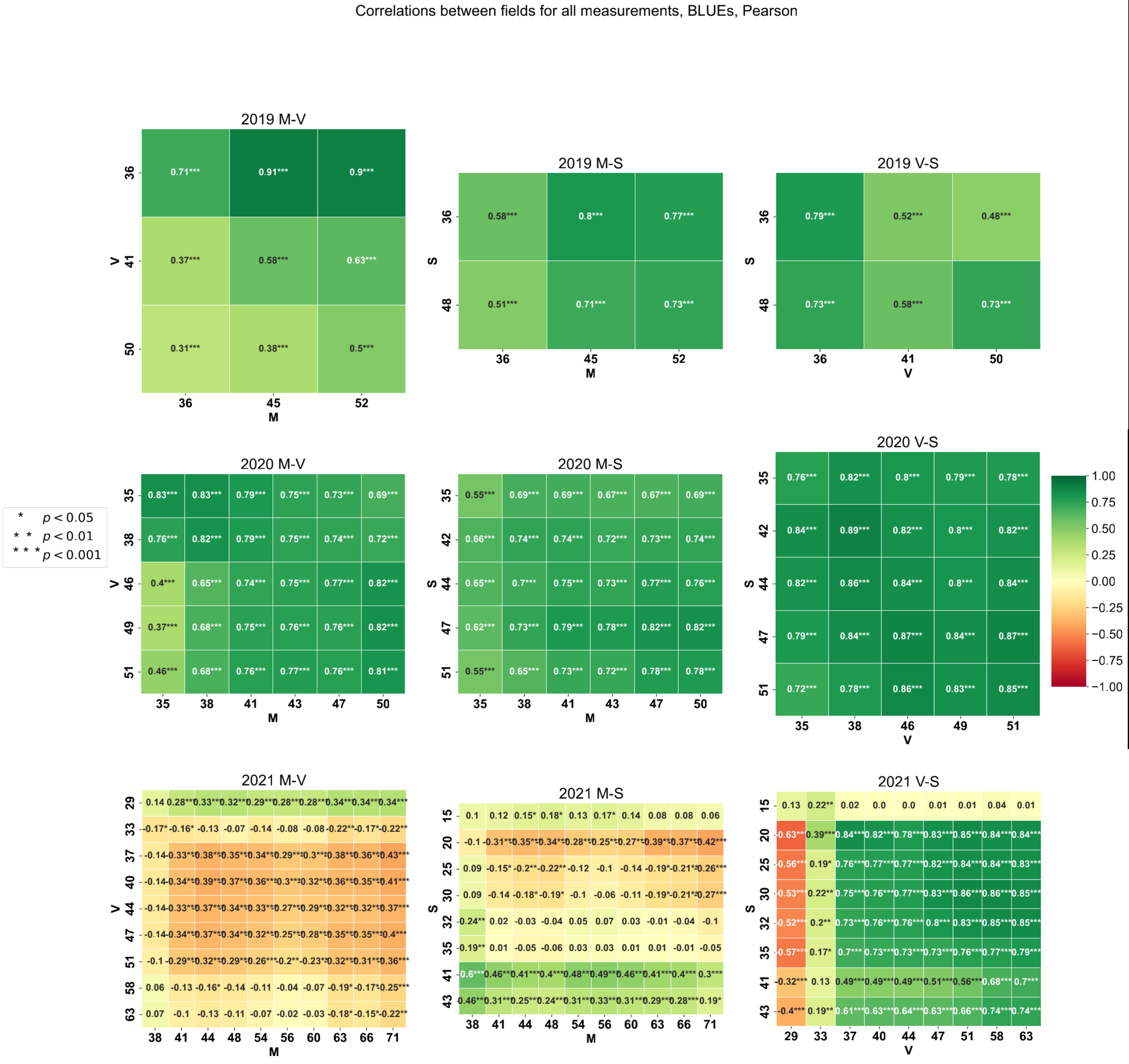}
    \caption{Pearson's correlation coefficient for the canopy sizes on different dates (DAPs) between the three test fields in each year.}
    \label{fig:all_p_corr}
\end{figure}

In Figure~\ref{fig:all_p_corr} one can see how the choice of the DAP affects the (Pearson) correlation in the vigor parameters between the test fields -- the quantity which is eventually displayed in Figure~\ref{fig:correlations}. It is clear that, due to the adverse weather conditions in France in 2021, no choice DAP could significantly improve the correlations with the test field M in 2021.

\begin{table}[]
    \centering
    \begin{tabular}{c|c|c|c|}
            & M & S & V\\
                \hline
       2019 & 52& 48& 36\\
       2020 & 47& 47& 51\\
       2021 & 54& 30& 58
    \end{tabular}
    \caption{Measurement dates (in DAPs) per test field and year, when the canopy area is considered to be the seedlot vigor parameter.}
    \label{tab:choice_dap}
\end{table}

\subsection*{Tuber data}

The seed lots were collected from the various seed growers and stored at 8 °C. From each seed lot 4 batches of 5 tubers were washed, dried and peeled. Both the flesh and peel fraction was cut into small pieces using ceramic knives and frozen in liquid nitrogen. Two 50 ml tubes were filled with frozen potato flesh and peel respectively, weighed and stored at -80 °C before freeze drying. A separate 150 ml container was filled with ~100 g frozen potato flesh, weighed and stored at -20 °C before freeze drying. All samples were weighed after freeze drying and dry matter percentages were calculated.

\subsubsection*{XRF data}
From the 150 ml container, 8.3 grams of freeze dried potato flesh was milled in a zirconium oxide grinder and sieved. A mixture of 8.0 g powder and 2.0 g binder was well homogenized and pressed into a pellet. The dry weight concentrations of Mg, K, Ca, Fe, Cu, Zn, P, S, Cl and Mn were determined using energy dispersive X-ray fluorescence spectroscopy (Bruker S2 PUMA) and converted to concentrations in fresh weight using the dry matter percentage.

\subsubsection*{FTIR data}
The freeze dried potato flesh of the 50 ml tube was ground to a fine powder and homogenized. The powder of each batch was put into 3 wells of a 96-well sample plate and measured in a high-throughput FTIR spectrometer (Bruker Tensor II + HTS-XT). Spectral outliers among the replicates were identified and the corresponding samples were remeasured. 

\subsubsection*{HSI data}
From each seed lot, ~50 tubers were selected and a 1 cm wide longitudinal slice from the center of each tuber was cut. The slices were placed on a moving platform and scanned using a push broom SWIR hyperspectral camera (900 -- 2500 nm, Specim). Before each scan, a dark and white reference was scanned and used to convert the data to absorbance units. 

\begin{figure}[h]
\begin{minipage}[t]{0.65\linewidth}
\hspace{0.8cm}
        \includegraphics[width=0.95\linewidth, trim={0cm 0.2cm 0cm 1.2cm},clip]{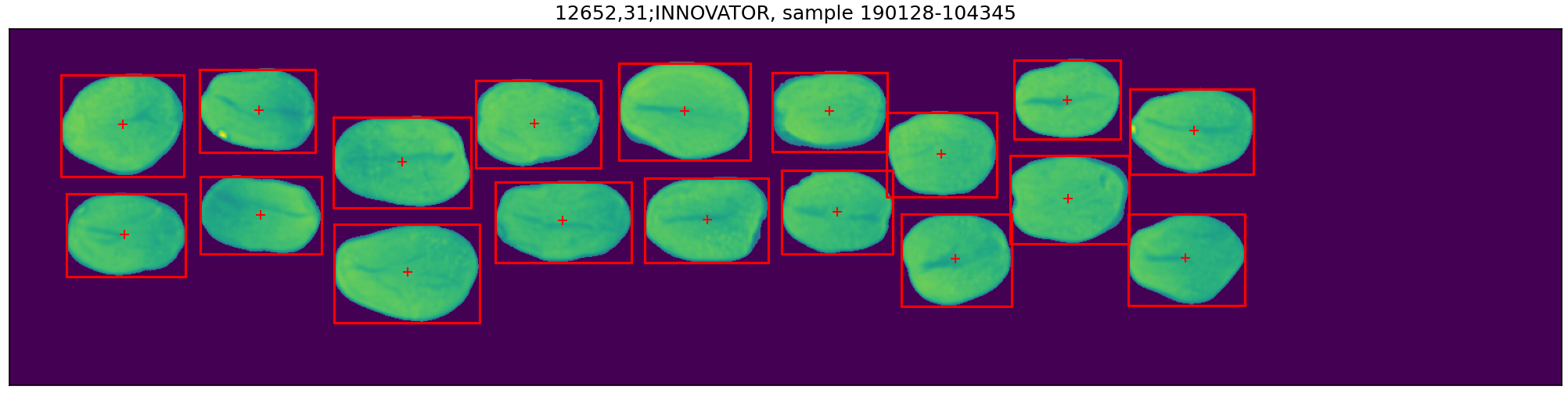}
\end{minipage}
\hspace{0.8cm}
\begin{minipage}[t]{0.3\linewidth}
        \includegraphics[height=0.48\linewidth, trim={10.2cm 15.19cm 8cm 1.2cm},clip]{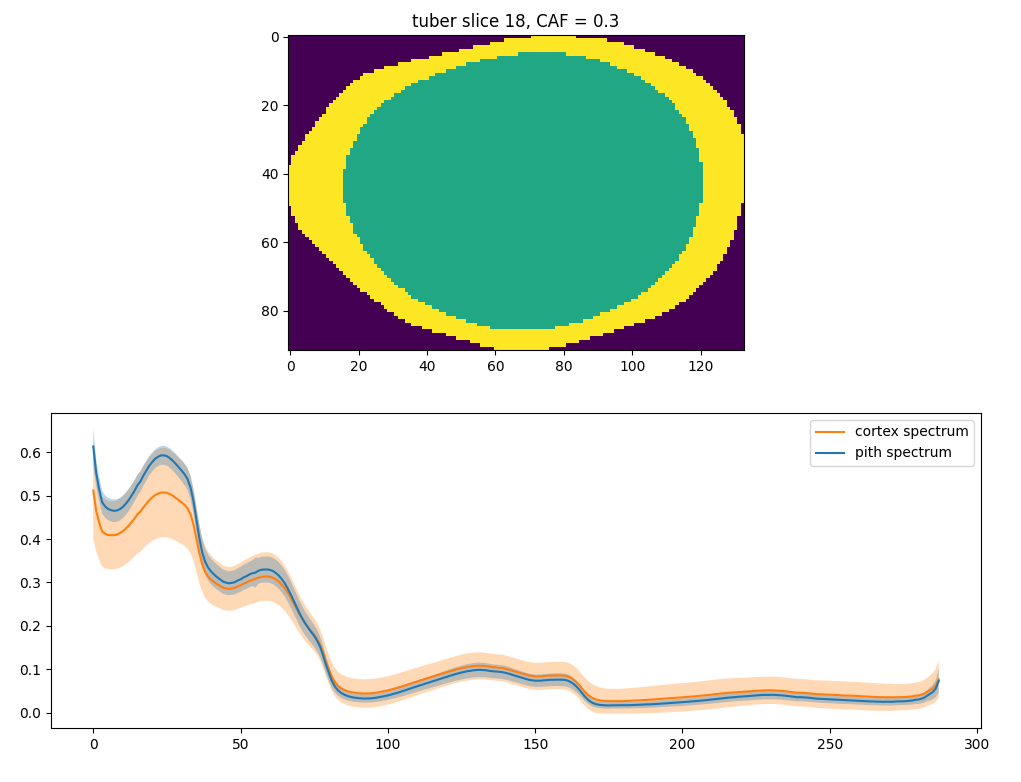}
\end{minipage}

        \includegraphics[width=0.95\linewidth, trim={0cm 0cm 0cm 14cm},clip]{Images/HSI-extraction/HSI-spectra-extraction.png}

    \caption{Extraction of the HSI spectral signatures. Top-left panel: spectrum-average image of the tray containing the seedlot tuber slices with the bounding boxes of individual slices. Top-right panel: the result of the segmentation of a single slice into its cortex (outer, yellow) and pith (inner, green) parts. Bottom panel: the spectra of the all cortex and pith pixels (shaded) and the corresponding average spectral signatures.}
    \label{fig:HSI-extraction}
\end{figure}

For each seedlot, a hyper-spectral image containing a number of tuber slices was acquired, see Figure~\ref{fig:HSI-extraction} (upper-left image). The sub-image of each individual slice was automatically detected and segmented into two biologically meaningful partitions: pith and cortex. The average spectral signature per partition was calculated as the mean of spectra for all pixels of that part. Then, the average of the spectral signatures over all slices in the image was taken. Thus, the HSI seedlot data represent the average spectral signatures of the cortex and pith of a seedlot, see Figure~\ref{fig:HSI-extraction} (bottom plot).

Technically, the most involved stage of this process is the automated segmentation of the slices into the pith and cortex parts. The outer boundary of each slice was easy to detect on the gray-scale image obtained by averaging over the spectral dimension, see Figure~\ref{fig:HSI-extraction} (upper-left image). Geometrically, the cortex part was defined as the domain of the slice bounded on the outside by the outer boundary of the slice and on the inside by the level curve of the potential function that solved the two-dimensional Poisson equation with the point source at the geometrical center (centroid) of the slice and zero Dirichlet boundary conditions at the outer boundary of the slice. The main property of this level curve is that it conforms to the outer boundary for the zeroth level of the potential and gradually changes its shape towards a circle around the point source located at the centroid for higher levels of the potential. For each slice, the level of the curve was chosen in such a way that the area of the cortex part would constitute 30\% of the total slice area, see Figure~\ref{fig:HSI-extraction} (upper-right image).

\subsubsection*{Metabolome}

From both sample types (flesh and peel) untargeted metabolic profiles were measured in an incomplete block design following guidelines from the Metabolomics Quality Assurance and Quality Control Consortium (mQACC) \cite{metabolomics}. Pooled samples were created by mixing 60 random lots, 10 per variety, for flesh and peel separately. These samples were used in separate dilution series and as precision references every 10 sample injections in the experiment runs.

The freeze dried sample (100 mg) was weighed in a 2 ml Eppendorf tube, 1.3 ml of methanol was added, vortexed and shaken for 30 minutes. After centrifugation (14000 rpm, 5 min.), 1 ml supernatant was transferred to a new tube and dried using a vacuum concentrator (1 mbar, 35 °C,150 min.). The residue was redissolved in cyclohexane (200 µl), milliQ water (300 µl) was added, shaken (10 min.) and centrifuged (14000 rpm, 10 min.). Using extended length tips, 180 µl of the lower aqueous phase was transferred to a 0.2 µl PVDF filter plate and centrifuged (1200 rpm, 4 min.). 

The filtered samples were analyzed using a Waters Acquity I-class UPLC coupled with a Waters Xevo G2-XS QTOF MS. Chromatographic separation was achieved on a reverse-phase Acquity UPLC HSS T1.8 µm (2.1 x 100 mm) column (Waters) at 40°C. The mobile phases employed were A: water (MilliQ) containing 0.1\% formic acid (UPLC grade, BioSolve) and B: acetonitrile (UPLC grade, BioSolve). The flow rate was maintained at 0.5 mL/min, and the injection volume was 5 µL. A gradient elution was employed, starting with 99\% A for the first minute, followed by a linear gradient to 30\% A over the next 10 minutes. A cleanup step was performed at 99\% B for 1 minute, followed by 3 minutes of re-equilibration at the initial conditions. The total runtime was 15 minutes. The ESI source was operated in both positive and negative ionization modes with the following settings: capillary voltage, 3.00 kV; cone voltage, 40 V; source temperature, 120°C; desolvation temperature, 350°C; gas flow rate, 800 L/h (N2); cone gas flow rate, 50 L/h (N2). Leucine enkephalin was used as a Lock Spray reference.

Mass data were collected over the m/z range of 50-1200. MSe (Multiplexed Selected Ion Monitoring) was employed to acquire both parent ions and fragmentation data in a single run. The collision energy ramp was set from 10 to 40 V, allowing for the fragmentation of precursor ions across a range of energies.

Alignment of the chromatograms and peak picking was performed in Progenesis QI. The precision reference results were used to perform batch and drift correction for each individual peak in the nPYc-Toolbox \cite{nPYc-Toolbox}. Feature selection was done based on the linearity in the dilution series and relative standard deviation in the precision references.

\subsection*{Multiple Linear Regression using Discontinuous-Galerkin Inverse Regression Model}

The explanatory and predictive model of the potato plant vigor in terms of the seed tuber properties combines diverse datasets with some of them, like the metabolome data, potentially causing the so called zero inflation. This happens when a certain metabolite is detected only in a few samples and is absent in the remaining samples. Other datasets, like FTIR and HIS, where each feature is the absorbance at a given frequency, are generally continuous not only over the frequency range but also across the samples. Here we employ the Inverse Regression (IR) model \cite{Atza-Budko-2024} that accommodates for this potential diversity in the predictor data via the freedom to choose a suitable functional basis to represent the predictor variables $x_{j}$, $j=1,\dots p$ as functions of the dependent variable $y$. Due to the aforementioned possibility of discontinuities in the predictor data across the samples, we apply the Discontinous Galerkin Finite-Element Method (DG-FEM) with the first-order (linear) Lagrange basis functions \cite{DG-FEM-book}. Since this is the first time the FEM is applied in the multiple linear regression and possibly also in the general Machine-Learning context, this section elaborates the technical details of the approach.

Let $X\in{\mathbb R}^{n\times p}$ be the predictor data matrix and ${\bm y}\in{\mathbb R}^{n}$ -- the vector of vigor data sorted by the magnitude of its entries. The idea of the IR method is to represent the predictor data as $x_{i,j}=x_{j}(y_{i})$, i.e., as the (sampled) functions of the dependent variable $y$. These functions are assumed to belong to a finite-dimensional function space defined by the choice of a suitable basis, i.e., a finite-dimensional set of basis functions. With the FTIR data \cite{Atza-Budko-2024} we have applied the polynomial basis and a very simple projection procedure leading to the decomposition $X=VA$ of the data matrix in terms of the basis matrix $V$ and the coefficient matrix $A$. A more rigorous and general projection procedure, suitable for a wider class of functions, is the Galerkin projection scheme widely used in the numerical solution of partial differential equations \cite{FEM-book}.

The derivation of the DG-FEM formulation of the IR model starts from the discretization of the $y$-range, i.e., the interval $I=\left[min(\bm y) , max(\bm y)\right]$, into $m$ finite elements $I_{k}$. The number of elements $m$ plays the role of a regularization parameter that will eventually be chosen through a cross validation (CV) procedure, similarly to the maximal polynomial degree in the polynomial-basis IR method \cite{Atza-Budko-2024} or the maximal dimension of the Krylov subspace in the PLS method \cite{ChemometMLR2001}. This $y$-range discretization yields $m+1$ delimiters $min(\bm y)=y_0, y_1, \dots, y_{m-1}, y_m=max(\bm y)$. Here, for simplicity, we consider the uniform mesh with the step size $h=y_{k+1}-y_k$.

Next, on each element $I_k= \left[y_k, y_{k+1}\right]$, we introduce two local linear Lagrange basis functions $v_{k,1}(y)$ and $v_{k,2}(y)$, defined as:
\begin{align}
      v_{k,1} (y)=\begin{cases}
    \frac{y_{k+1}-y}{h}, & \text{if $y \in I_k$}\\
    0, & \text{otherwise}
  \end{cases},
\;\;\;\;\;\;\;\;
v_{k,2} (y)=\begin{cases}
    \frac{y- y_{k}}{h}, & \text{if $y \in I_k$}\\
    0, & \text{otherwise}
  \end{cases}.
\end{align}

The main assumption of the IR method is that the predictor data are `generated' by the functions $x_{j}(y)$, $j=1,\dots,p$; all of which can be represented in terms of the chosen basis as:
\begin{align}
    x_j(y)=\sum_{k=1}^{m}\sum_{q=1}^2 \alpha^{(j)}_{k,q} v_{k,q}(y) 
    \label{eq: continuous-repr},
\end{align}
and the observed predictor data $x_{i,j}$ are simply the sampled values $x_{j}(y_{i})$, $j=1,\dots,p$; $i=1,\dots,n$.

The goal of the model training is to determine (learn) the expansion coefficients $\alpha_{k,q}, \; k=1, \dots, m, \; q=1,2$ in the representation (\ref{eq: continuous-repr}). For this purpose, we apply the Galerkin scheme, where one chooses the {\it test} functions to be equal to the basis functions. Multiplying both sides of the equation (\ref{eq: continuous-repr}) with a test function and integrating over the $y$-range, we arrive at the $2m$ equations for the $2m$ unknown coefficients $\alpha_{k,q}$:
\begin{align}
     \int_{y_0}^{y_m} x_j(y) v_{\hat{k}, \hat{q}}(y)\; dy = \sum_{k=1}^{m}\sum_{q=1}^2 \alpha^{(j)}_{k,q} 
     \int_{y_0}^{y_m}  v_{k, q}(y) v_{\hat{k}, \hat{q}}(y)\; dy,\;\;\;\;\;\;\hat{k}=1, \dots, m; \; \hat{q}=1,2.
\end{align}
To further specify the system matrix and the right-hand side vector of this linear algebraic system of equations, we first note that, by the nature of the DG finite-element basis, the following partial orthogonality condition holds:
 \begin{align}
 \int_{y_0}^{y_m}  v_{k, q}(y) v_{\hat{k}, \hat{q}}(y)\; dy =\delta_{k, \hat{k}} \int_{y_0}^{y_m}  v_{\hat{k}, q}(y) v_{\hat{k}, \hat{q}}(y)\; dy.
 \end{align}
Therefore, the equations for the coefficients $\alpha_{k,q}$ corresponding to the different elements $I_{k}$ decouple, and it is possible to find all coefficients by solving $m$ separate $2\times 2$ linear systems:
\begin{align}
      \int_{y_{\hat{k}-1}}^{y_{\hat{k}}} x_j(y) v_{\hat{k}, \hat{q}}(y)\; dy = \sum_{q=1}^2 \alpha^{(j)}_{\hat{k},q} \int_{y_{\hat{k}-1}}^{y_{\hat{k}}}  v_{\hat{k}, q}(y) v_{\hat{k}, \hat{q}}(y)\; dy,\quad \hat{k}=1,\dots, m; \; \hat{q}=1,2. \label{eq:continuous-integrals}
\end{align}

The integrals in (\ref{eq:continuous-integrals}) are further approximated by a numerical quadrature rule \cite{numerical-book}. Since the values of $x_j(y)$ are only given at the available data points $y_1, \dots, y_n$, the trapezoidal rule seems to be appropriate:
\begin{align}
\label{eq: trapesoidal}
\begin{split}
\int_{y_{\hat{k}-1}}^{y_{\hat{k}}} v_{\hat{k}, q}(y) v_{\hat{k}, \hat{q}}(y)\; dy &= \sum_{m=1}^{n_{\hat{k}}} w_{\hat{k},m} v_{\hat{k}, q}(y_{\hat{k},m})v_{\hat{k}, \hat{q}}(y_{\hat{k},m}) +{\mathcal O}(\max_{m}\vert y_{\hat{k},m+1}-y_{\hat{k},m}\vert^{2}),
\\
\int_{y_{\hat{k}-1}}^{y_{\hat{k}}} x_j(y) v_{\hat{k}, \hat{q}}(y)\; dy &= \sum_{m=1}^{n_{\hat{k}}} w_{\hat{k},m} x_j(y_{\hat{k},m})v_{\hat{k}, \hat{q}}(y_{\hat{k},m})+{\mathcal O}(\max_{m}\vert y_{\hat{k},m+1}-y_{\hat{k},m}\vert^{2}),
\end{split}
\end{align}
where the data points $y_{\hat{k},m}$, $m=1, \dots, n_{\hat{k}}$ are situated within the 
element $I_{\hat{k}}$ bounded by the delimiters $y_{\hat{k}}$ and $y_{\hat{k}+1}$. The weights $w_{\hat{k},m}$ of the trapezoidal quadrature can be explicitly written out as:
\begin{align}
\begin{split}
    w_{\hat{k},1}&= \vert y_{\hat{k},1} - y_{\hat{k}}\vert + \frac{1}{2} \vert y_{\hat{k},2}-y_{\hat{k},1}\vert,
    \\
    w_{\hat{k},m}&=\frac{1}{2} \vert y_{\hat{k},m}-y_{\hat{k},m-1}\vert+\frac{1}{2} \vert y_{\hat{k},m+1}-y_{\hat{k},m}\vert,\;\;\;m=2,\dots,n_{\hat{k}-1},
    \\
    w_{\hat{k},n_{\hat{k}}}&= \vert y_{\hat{k},n_{\hat{k}}} - y_{\hat{k}+1}\vert + \frac{1}{2} \vert y_{\hat{k},n_{\hat{k}}}-y_{\hat{k},n_{\hat{k}}-1}\vert.    
\end{split}
\end{align}

Therefore we can write a linear approximation of Equation \ref{eq:continuous-integrals} for a given interval $I_{\hat{k}}$ with the following notation:
\begin{equation}
    V_{\hat{k}}=\begin{bmatrix}
        v_{\hat{k},1}(y_{\hat{k} ,1}) &v_{\hat{k},2}(y_{\hat{k} ,1}) \\
        v_{\hat{k},1}(y_{\hat{k} ,2}) &v_{\hat{k},2}(y_{\hat{k} ,2}) \\
        \vdots&\vdots\\
        v_{\hat{k},1}(y_{\hat{k} ,n_{\hat{k}}}) &v_{\hat{k},2}(y_{\hat{k} ,n_{\hat{k}}}) \\
    \end{bmatrix}, 
    \quad 
    W_{\hat{k}}={\rm diag}\left(\begin{bmatrix}
        w_{\hat{k},1}\\
        \vdots\\
        w_{\hat{k},n_{\hat{k}}}
    \end{bmatrix}\right), 
    \quad 
    \bm{x}_j=\begin{bmatrix}
    x_j(y_{\hat{k} ,1})\\
    \vdots\\
    x_j(y_{\hat{k} ,n_{\hat{k}}})
    \end{bmatrix}
\end{equation}
as
\begin{equation}
    V_{\hat{k}}^TW_{\hat{k}}\bm{x}_j= V_{\hat{k}}^TW_{\hat{k}}V_{\hat{k}}\begin{bmatrix}
        \alpha_{\hat{k}, 1}^{(j)} \\   \alpha_{\hat{k}, 2}^{(j)}
    \end{bmatrix} = V_{\hat{k}}^TW_{\hat{k}}V_{\hat{k}} \bm{\alpha}_j
\end{equation}
where the column vector $\bm{\alpha}_j$ is a subvector of $\bm{a}_j$, the j-th column of $A$, relative to the $\hat{k}$-th interval.

Since the linear systems concerning each interval are decoupled, the full linear system is:
\begin{equation}
     V^TWX= V^TWV A, \quad V^TW\bm{y}= V^TWV\bm{a}_0\label{eq:full_lin_sys}
\end{equation}
where
\begin{equation}
    V=\begin{bmatrix}
        V_1 & &&\\
        &\ddots&&\\
        &&V_k&\\
        &&&\ddots\\
        & &&& V_m
    \end{bmatrix}, \quad \quad     W=\begin{bmatrix}
        W_1 & &&\\
        &\ddots&&\\
        &&W_k&\\
        &&&\ddots\\
        & &&& W_m
    \end{bmatrix}, \quad \quad \bm{a}_j=\begin{bmatrix}
         \alpha_{1, 1}^{(j)} \\   
         \alpha_{1, 2}^{(j)}\\
         \vdots\\
        \alpha_{m, 1}^{(j)} \\   
         \alpha_{m, 2}^{(j)}
    \end{bmatrix}.
\end{equation}

Therefore the sought after matrix $A$ and vector $\bm{a}_0$ solving Equations \ref{eq:full_lin_sys} are:
\begin{equation}
    A=(V^TWV)^{-1}V^TWX
\end{equation}
and  $\hat{\bm{y}}_t$, the prediction of the vector $\bm{y}_t$:
\begin{align}
    \hat{\bm{y}}_t&=\hat{V}\bm{a}_0\\
    &=X_t A^T (AA^T)^{-T} (V^TWV)^{-1}V^TW\bm{y}_t
\end{align}
where $\hat{V}=X_t A^T (AA^T)^{-T}$. 

By denoting the full dataset as $S=\left[\bm{y} | X\right]$, and the test set as $S_t$, the predictions for both the dependent and independent variables can be written as:
\begin{equation}
        \hat{S}=S_t A^T (AA^T)^{-T}A.
\end{equation}

\subsection*{Measures of prediction quality}
The standard measures of the MLR model quality, such as the $R^{2}$ coefficient, are not applicable to the severely overparameterized case, where the number of predictors is significantly larger than the number of experiments \cite{Izenman2008}. In this paper, both the measured, ${\bm y}$, and the predicted, $\hat{\bm y}$, vigor data vectors, each of length $n$, are normalized as:
\begin{align}
\label{eq:normalization}
\tilde{\bm y} = \frac{1}{\left\Vert{\bm y}- n^{-1}{\bm 1}{\bm 1}^{T}{\bm y}\right\Vert_{2}}\left({\bm y} - n^{-1}{\bm 1}{\bm 1}^{T}{\bm y}\right),
\;\;\;\;\;\;
\tilde{\hat{\bm y}} = \frac{1}{\left\Vert\hat{\bm y}- n^{-1}{\bm 1}{\bm 1}^{T}\hat{\bm y}\right\Vert_{2}}\left(\hat{\bm y} - n^{-1}{\bm 1}{\bm 1}^{T}\hat{\bm y}\right).
\end{align}
Here, ${\bm 1}\in{\mathbb R}^{n}$ is the vector of all ones.
After this transformation, we have $\Vert\tilde{\bm y}\Vert_{2}=\Vert\tilde{\hat{\bm y}}\Vert_{2}=1$, where $\Vert\cdot\Vert_{2}$ denotes the Euclidean norm of the vector, and $n^{-1}{\bm 1}^{T}\tilde{\bm y}=n^{-1}{\bm 1}^{T}\tilde{\hat{\bm y}}=0$, i.e., both vectors have the unit norm and the zero mean.

For the normalized vectors we compute the squared norm of the residual vector, also known as Sum of Squared Errors (SSE):
\begin{align}
\label{eq:residual}
r^{2} = \Vert\tilde{\bm y}-\tilde{\hat{\bm y}}\Vert_{2}^{2},
\end{align}
which is a measure of the unexplained variance. In this case, the $r^{2}$ is related to the Pearson correlation coefficient $c(\tilde{\bm y},\tilde{\hat{\bm y}})$ between the measured and predicted data vectors as follows:
\begin{align}
\label{eq:pearson-r}
r^{2} = 2(1-c).
\end{align}
Since $-1\leq c\leq 1$, we have the bounds $0\leq r^{2}\leq 4$. We consider the model to be predictive if for the testing dataset it gives the values $r^{2}<1$ and $c>0.5$, with the latter parameter also subject to the $p$-value analysis. The upper bound $r^{2}=1$ for the SSE and the corresponding lower bound $c=0.5$ for the correlation, stem from the fact that for $1\leq r^{2}\leq 4$ the prediction $\tilde{\hat{\bm y}}$ is usually of very poor quality, making it impossible to correctly classify the vigor of the seedlot as belonging to the low, middle or high quantile.

\subsection*{Training, validation and testing}

Both the variety-agnostic and the six variety-specific DG-IR models contain two hyper-parameters that need to be tuned. The first is the number $m$ of segments, see the Equation~(\ref{eq: continuous-repr}), that divide the range of the vigor parameter. The more segments are used, the better is the fit for each predictor variable. The $10$-fold Cross Validation is used to find the optimal number of segments and avoid overfitting. In the DG-IR method with linear basis functions, at least two data points should be present within each segment. This puts a practical upper bound on the number of uniform segments, so that one only has to check for a smaller number of segments in search of the optimal parameter~$m$. The results presented in Figure~\ref{fig:best_col_fits} illustrate the fits with the optimal number of segments determined by the CV procedure.

The second hyper-parameter is the threshold $\tau$ which allows discarding the predictors with the $X$-data residual above $\tau$, see the dashed horizontal lines in Figure~\ref{fig:col-wise-residuals}. Tuning of this parameter is performed with the number of segments $m$ fixed at the optimal value.

To tune the above-mentioned hyper-parameters of the DG-IR model, the dataset is split into the training-validation and testing subsets. When the testing is performed on a different year, then the complete dataset of that year is used as the testing set. This means that the complete dataset of the training year can be used for training and validation (parameter tuning). When, however, the testing is performed on the same year as the training, then 33\% of the variety data is reserved for testing and the remaining 67\% is used for training and validation.

\section*{Code and Data availability}
Both the seed tuber and plant canopy datasets are available at \cite{dataset_this_paper}. The data in \cite{dataset_this_paper} also includes the Python code necessary to reproduce the results of regression.

\section*{Acknowledgements}

We are grateful to our industrial partners HZPC and Averis Seeds 
B.V. for collecting the seed tuber data and supporting the field trials. In particular we would like to thank 
Johan Hopman from Averis Seeds B.V for the many fruitful discussions and valuable insights.  

\section*{Funding}

This work was funded by HZPC Research B.V., Averis Seeds B.V., BO-Akkerbouw, and European Agricultural Fund for Rural Development.

\section*{Author contributions statement}
EA and NVB performed quantitative analysis of drone images, modeling and data analysis.
EA, NVB, and RK wrote the manuscript. 
FvdW, HvD, FH and RK coordinated sampling collection and experimental field trials. 
HZPC Research was responsible for all instrumental measurements.
FH and RK were responsible for data collection.
All authors reviewed the manuscript. 

\section*{Additional information}

\textbf{Competing interests} 
The authors declare that this study received funding from HZPC Research B.V. and Averis Seeds B.V. 
The funder had the following involvement in the study: study design, sample collection, and the decision to submit it for publication. RK, FH, FvdW, HvD are currently employed by HZPC research B.V.

\end{document}